\documentclass[acmsmall, authorversion=true, nonacm=true]{acmart}
\usepackage[T1]{fontenc}
\usepackage{placeins}
\usepackage{amsfonts}
\usepackage{comment}
\usepackage{graphicx}
\usepackage{enumitem}
\usepackage{stfloats}
\usepackage[detect-none]{siunitx}
\usepackage{float}
\usepackage{xcolor}
\usepackage[noend]{algpseudocode}
\usepackage{algorithm}
\usepackage{wrapfig}
\usepackage{subcaption}
\usepackage{caption}
\usepackage{booktabs}
\usepackage{xspace}
\usepackage{tikz}
\usepackage{setspace}
\usepackage{lmodern}

\usepackage{xhfill}
\usepackage{dashrule}
\usepackage[capitalise]{cleveref}
\crefname{section}{Sect.}{Sects.}
\Crefname{section}{Section}{Sections}
\crefname{algorithm}{Alg.}{Algs.}
\Crefname{algorithm}{Algorithm}{Algorithms}
\crefname{line}{line}{lines}
\usepackage{scalerel}
\usepackage{relsize}
\usepackage{mathtools}

\usepackage{multibib}

\newcommand{\Data}[1]{\hspace*{\algorithmicindent}\hspace*{0pt}\textbf{Data:} \hspace*{5.8pt}#1\\}
\newcommand{\Datan}[1]{\hspace*{\algorithmicindent}\hspace*{30.8pt}#1\\}
\newcommand{\Global}[1]{\hspace*{\algorithmicindent}\hspace*{0pt}\textbf{Global:} \hspace*{-1pt}#1\\}
\newcommand{\Globaln}[1]{\hspace*{\algorithmicindent}\hspace*{30.8pt}#1\\}
\newcommand{\Param}[1]{\hspace*{\algorithmicindent}\hspace*{0pt}\textbf{Inputs:} #1\\}
\newcommand{\Paramn}[1]{\hspace*{\algorithmicindent}\hspace*{30.8pt}#1}

\newcommand{\just}[1]{\ensuremath{\mathit{#1}}}

\newcommand{\cmod}{\ \textbf{mod}\ }

\newcommand{\Cpp}{C\hspace{-.05em}\texttt{+\hspace{-.05em}+}\xspace}
\newcommand{\Cc}{C\xspace}
\newcommand{\bgrate}{\just{background\_dirty\_ratio}\xspace}
\newcommand{\dirtyratio}{\just{dirty\_ratio}\xspace}
\newcommand{\dirtyrate}{\emph{dirty rate}\xspace}
\newcommand{\setpoint}{\just{setpoint}\xspace}
\newcommand{\posratio}{\just{pos\_ratio}\xspace}
\newcommand{\dirtyexpire}{\just{dirty\_expire}\xspace}
\newcommand{\mbw}{\just{bw_{mem}}\xspace}
\newcommand{\rambw}{\just{bw_{\emph{ramdisk}}}\xspace}
\newcommand{\sbw}{\just{bw_{dev}}\xspace}

\newcommand{\srbw}{\just{bw_{rdev}}\xspace}
\newcommand{\syswcost}{\just{sc_w}\xspace}
\newcommand{\sysswcost}{\just{sc_{sw}}\xspace}
\newcommand{\skcost}{\just{c_{sk}}\xspace}
\newcommand{\israndom}{\just{is\_rnd}\xspace}
\newcommand{\bs}{\just{bs}\xspace}
\newcommand{\cbf}{\just{bf}\xspace}
\newcommand{\isfit}{\just{is\_fit}\xspace}

\newcommand{\syncti}{\just{t_{i}^{sync}}\xspace}

\newcommand{\nextsysbw}{Syscall\_IO\_Cost}
\newcommand{\nextlibbw}{Library\_IO\_Cost}
\newcommand{\sizes}{\just{sizes}}
\newcommand{\delays}{\just{delays}}

\newcommand{\bglimit}{\just{limit_{bg}}}
\newcommand{\hlimit}{\just{limit_{hard}}}
\newcommand{\dirty}{\just{dirty}}
\newcommand{\bw}{\just{taskrate}}
\newcommand{\avgbw}{\just{bw_{avg}}}
\newcommand{\ttime}{\just{time}}

\newcommand{\iocost}{\just{cost}}
\newcommand{\iotime}{\just{io\_time}}

\newcommand{\haveexpired}{\just{exp}}
\newcommand{\bgflush}[1]{\just{background\_flush(#1)}}

\newcommand{\freerun}{\just{free\_run}}
\newcommand{\asyncrun}{\just{async\_run}}
\newcommand{\throttlerun}{\just{throttle\_run}}
\newcommand{\period}{\just{interval}}

\newcommand{\size}{\just{size}}
\newcommand{\delay}{\just{delay}}

\newcommand{\pending}{\just{pending}}
\newcommand{\pendingd}{\just{delay_{acc}}}
\newcommand{\rest}{\just{rest}}
\newcommand{\rem}{\just{rem}}

\newcommand{\datablock}{\just{datablock}}
\newcommand{\data}{\just{p}}
\newcommand{\offset}{\just{offset}}
\newcommand{\file}{\just{file}}
\newcommand{\activedata}{\just{active}}
\newcommand{\ioendtime}{\just{endtime}}
\newcommand{\pagecache}{\just{cache}}

\newcommand{\updatecache}[1]{\just{update\_cache(#1)}}
\newcommand{\balancecache}[1]{\just{balance\_cache(#1)}}
\newcommand{\tobecleaned}{\just{evictable}}

\newcommand{\insertlist}[1]{\just{put(#1)}}
\newcommand{\matchingblocks}{\just{matching\_blocks}}
\newcommand{\block}{\just{b_i}}
\newcommand{\cachedisj}{\just{disjuncts_{cache}}}
\newcommand{\newdisj}{\just{disjuncts_{new}}}

\newcommand{\redbw}{\just{bw_{reduced}}}
\newcommand{\newdblock}{\just{newblock}}
\newcommand{\nextlim}{\just{limit}}
\newcommand{\limblock}{\just{cut}}
\newcommand{\overlap}{\just{overlap}}

\newcommand{\overlapfn}[1]{\just{overlapping\_blocks(#1)}}
\newcommand{\cutblock}[1]{\just{cut\_block(#1)}}
\newcommand{\prevpos}{\just{prev\_pos}}
\newcommand{\prevoffset}{\just{offset_{bf}}}

\newcommand{\expired}{$( \,\exists \data\in\pagecache:     \data\just{.\ioendtime} < \ttime - \dirtyexpire \,)$}

\newcommand{\func}[1]{\fontfamily{lmtt}\fontseries{m}\selectfont{\texttt{#1}}}

\newcommand{\KB}{\mathrm{KB}}
\newcommand{\MB}{\mathrm{MB}}
\newcommand{\GB}{\mathrm{GB}}

\newcommand{\OSYNC}{{\smaller O\_SYNC}\xspace}
\newcommand{\ODSYNC}{{\smaller O\_DSYNC}\xspace}
\newcommand{\ODIRECT}{{\smaller O\_DIRECT}\xspace}

\definecolor{doccolor}{gray}{0.47}

\newcommand{\comdoc}[1]{\textcolor{doccolor}{\hfill\,\,$\triangleright$\,\,\textit{#1}}}

\algnewcommand{\IfThenElse}[3]{%
\algorithmicif\ #1\ \algorithmicthen\ #2\ \State \algorithmicelse\ #3}

\newcommand{\pluseq}{\mathrel{+}=}
\newcommand{\minuseq}{\mathrel{-}=}

\usepackage{fontawesome5} 
\newcommand{\orcidid}[1]{\href{https://orcid.org/#1}{\textcolor[HTML]{A6CE39}{$^{\textrm{\faOrcid}}$}}}

\title{CAWL: A Cache-aware Write Performance Model of Linux Systems}

\author{Masoud Gholami}
\affiliation{\orcidid{0000-0001-5331-6883}\institution{Zuse Institute Berlin}\city{Berlin}\country{Germany}}

\author{Florian Schintke}
\affiliation{\orcidid{0000-0003-4548-788X}\institution{Zuse Institute Berlin}\city{Berlin}\country{Germany}}

\begin{document}

\begin{abstract}
The performance of data intensive applications is often dominated by
their input/output (I/O) operations but the I/O stack of systems is
complex and severely depends on system specific settings and hardware
components.
This situation makes generic performance optimisation challenging and
costly for developers as they would have to run their application on a
large variety of systems to evaluate their improvements.
Here, simulation frameworks can help reducing the experimental
overhead but they typically handle the topic of I/O rather
coarse-grained, which leads to significant inaccuracies in performance
predictions. Here, we propose a more accurate model of the write
performance of Linux-based systems that takes different I/O methods
and levels (via system calls, library calls, direct or indirect,
etc.), the page cache, background writing, and the I/O throttling
capabilities of the Linux kernel into account.
With our model, we reduce, for example, the relative prediction error
compared to a standard I/O model included in SimGrid for a random I/O
scenario from 67\,\% down to 10\,\% relative error against real
measurements of the simulated workload. In other scenarios the
differences are even more pronounced.
\end{abstract}



\maketitle

\section{Introduction} \label{sec:intro}
    Simulation and modeling of processes and workflows is a common
    technique to study and predict the performance characteristics of
    applications without the need to actually execute them with all
    studied parameter settings. Two exemplary popular frameworks for
    estimating the costs of running HPC applications and workflows are
    SimGrid~\cite{simgrid} and WRENCH~\cite{wrench}. They focus on
    modeling computation time and process interdependencies caused by
    network traffic and message exchange.
    Another important aspect for the application's performance is the
    input/output (abbreviated as I/O or even IO in the following) they
    perform, which SimGrid and WRENCH address on a coarse grain level.
    But for a realistic model of the IO behaviour, additional aspects
    of the IO subsystems, such as different buffers and IO bandwidth
    steering over time, play a vital role as they influence the
    observed IO costs. We provide accurate models to estimate the IO
    costs considering the cache involvement of the OS and \Cc standard
    library. Such models allow improving the accuracy of HPC and
    workflow simulation frameworks for IO intensive applications.

    Following an overview of the main available IO methods of Linux
    and the \Cc standard library (\cref{sec:backg}) and a summary of
    the inner working of the Linux page cache mechanism
    (\cref{sec:linuxpagecache}), we present our
    main original contributions:
    \begin{itemize}
    \item We define models for the accurate prediction of the IO costs
      (write performance) for scenarios with and without the page
      cache and the buffer of the \Cc standard library
      (\cref{sec:model}). We distinguish:
      \begin{itemize}
      \item synchronized, direct IO with system calls (\cref{sec:mod-sync-direct-syscall}),
      \item synchronized, non-direct IO with system calls (\cref{sec:mod-sync-nondirect-syscall}),
      \item non-synchronized IO with system calls (\cref{sec:mod-nonsync}), and
      \item standard-library IO (\cref{sec:mod-libcost}).
      \end{itemize}
    \item We discuss related work and studies on IO modeling and storage
        devices in \cref{sec:rel}.
    \item We evaluate the IO cost prediction accuracy for
      various scenarios on different hosts and IO devices compared to
      measured system performance
      (\cref{sec:eval}):
    \begin{itemize}
    \item Here, we compare CAWL estimations (our contribution) with SimGrid
      IO estimates and show the importance and relevance
      that considering caching effects may have. Indeed, we observe the SimGrid
      IO model overestimating IO costs by up to $350\,\%$ and
      underestimating them by up to $99\,\%$ in different scenarios.
    \item In contrast, we show that CAWL estimates the actual IO costs
      mostly with more than $80-90\,\%$ accuracy in different cases.
    \end{itemize}
    \end{itemize}

\section{Background: File Input/Output Operations} \label{sec:backg}

    Programmatically, there are various options to read/write data from/to
    a file in \Cc/\Cpp. From the primitive file operations (i.e.,
    system calls or short syscalls)
    at the low level (e.g., $\func{open}$, $\func{write}$,
    etc.)~\cite[Ch.~2]{love2013} to the higher-level standard library functions
    such as $\func{fopen}$, $\func{fwrite}$, etc. of libc~\cite{libc},\cite[Ch.~3]{love2013},
    and the $\func{fstream}$ family of classes in the \Cpp standard
    library~\cite{stdlib}.
    To improve the IO performance of applications, the Linux kernel~\cite{linux},
    as well as the \Cc/\Cpp standard libraries, provide
    different caching mechanisms to buffer the IO in the main memory before
    they end up in the slower storage device.

    The Linux kernel
    defines a dynamic part of the
    main memory in the kernel space~\cite{kernelspace} as the
    \emph{page cache}~\cite[Ch.~2]{love2013},\cite[Sect.~16]{love2010linux_devel}.
    When a process issues a write operation, the Operating System (OS)
    copies the data to the page cache instead of directly writing it to the
    storage device (sync). Later, the OS will sync the data in the page
    cache with the backing storage device transparent to the application.
    On top of the page cache, the \Cc/\Cpp standard libraries provide
    a separate buffer in the user space for file IO.

    Caching the data instead of directly transferring them to the storage devices
    allows a quick return of the write operation, and
    consecutive reads do not need to engage the
    backing device for recently written data still in the
    cache. Additionally, the cache can help aggregate fine-granular random data accesses in small chunks to large data chunks that need fewer operations on the storage device. Moreover, highly-frequent modifications of
    the same data only affect the fast page cache in the main memory
    at first. The eventual state is written to the storage device
    delayed and more seldom.

    \subsection{Primitive File Operations} \label{sec:syscall}
    Linux offers, among others, the following primitive system calls
    for IO (see the respective manual page~\cite{man2} for
    details):

    \begin{itemize}
        \item $\func{open}$ opens a file and assigns
            a non-negative integer file descriptor for reference
            in other system calls.

        \item $\func{lseek}$ repositions the offset of an open
            file for read/write operations.

        \item $\func{write}$ receives the
            data to be written as input and
            copies the data into the
            \emph{page cache} (in the normal case of non-direct IO).

        \item $\func{fsync}$ receives a file descriptor as
            input and syncs its modified data and metadata from the page cache to the
            underlying storage device. Similarly, giving the \OSYNC~\cite[open]{man2} flag to
            the $\func{open}$ system call causes the written data to be synced
            before a $\func{write}$ returns.
            Calling $\func{fsync}$ is not necessary in most
            cases since the OS will eventually write the page cache's data
            to the storage device, transparent to the application (see
            \cref{sec:linuxpagecache}).

        \item $\func{read}$ reads data from a file. The OS
            first queries the page cache for the data. If not found, the
            data is fetched from the storage device.

        \item $\func{close}$ closes a file descriptor.
    \end{itemize}

    \subsection{Standard Library Functions} \label{sec:libcall}

    On top of the primitive file operations, which allow precise
    control over the IO operations, high-level standard functions are
    built, aiming for better portability instead. They are part of the \Cc
    standard library (see the respective manual page~\cite{man3} for
    details):

    \begin{itemize}
        \item $\func{fopen}$ opens a file as a stream and
            returns a $\texttt{FILE}$ structure pointer.

        \item $\func{fseek}$ is similar to $\func{lseek}$
            but works with $\texttt{FILE}$ pointers
            instead of integer file descriptors.

        \item $\func{fwrite}$ receives the data to be written
            as input and copies it into a buffer of several
            kilobytes (if it fits, see \cref{sec:mod-libcost}).
            If the data is larger than the buffer, $\func{fwrite}$
            invokes the $\func{write}$ system call to transfer the data
            to the page cache (details in \cref{sec:mod-libcost}).
            The buffer lies in the \emph{user space} portion of the
            main memory, as opposed to the \emph{page cache} that lies in the
            \emph{kernel space}.
            This way,
            frequent expensive system calls for every small data
            chunk are reduced to only one system call
            that writes data of relatively large size (several kilobytes).
            The $\func{fwrite}$ function returns as soon as it copied the data
            into its buffer. Only after the buffer is full (or when $\func{fseek}$
            or $\func{fflush}$ are called), $\func{fwrite}$ invokes the
            $\func{write}$ system call (\cref{sec:syscall}).
            As a result, the OS moves the data from the buffer
            to the page cache in the kernel space.
            Hence, the data goes through two
            layers of buffers (user space and kernel space).
            We discuss this behavior
            further in \cref{sec:mod-libcost}.

        \item $\func{fflush}$ invokes a $\func{write}$ system
            call to move the data from the library buffer into the page cache of
            the OS (into the kernel space). The data will still reside
            in main
            memory and is not necessarily synced to the disk. The OS manages the
            rest, as discussed in \cref{sec:syscall,sec:linuxpagecache}.

        \item $\func{fread}$ looks for the data first in the
            library buffer (user space). If found, it returns the data without
            issuing a system call. Otherwise, it invokes a $\func{read}$ system
            call (\cref{sec:syscall}).

        \item $\func{fclose}$ closes a file represented by a
            $\texttt{FILE}$ structure pointer.
    \end{itemize}

\section{Linux Page Cache Management} \label{sec:linuxpagecache}

After issuing a $\func{write}$ system call, the OS copies the data to
the page cache. Later, the OS will sync the data in the page cache
with the backing storage device in the background. Modified data still
not synced from the page cache to the disk is denoted as \emph{dirty}
data (\emph{dirty pages}). Linux defines the \emph{dirtyable} memory
as all the memory pages that processes can make dirty. Since Linux
3.14, \emph{dirtyable} memory consists of \textbf{free memory}, the
already \textbf{dirty pages}, and the \textbf{reclaimable} pages,
excluding the \textbf{reserved pages} of the main
memory~\cite{dirtyable,dirtylimits}. The OS tracks the \dirtyrate,
which is the \emph{dirty size} divided by the size of the
\emph{dirtyable} memory~\cite{dirtylimits}.
The Linux kernel manages the page cache by maintaining the \dirtyrate
between two limits: \bgrate and \dirtyratio~\cite{dirtylimits}. These
limits can be adjusted using the Linux command
\texttt{sysctl}~\cite{sysctl}. When the \dirtyrate reaches the
\bgrate, the kernel starts syncing the dirty pages in the background.
The \dirtyratio is treated as a hard limit for the \dirtyrate, which
should not be exceeded. Throughout this paper, we denote a process
that produces dirty pages (by issuing write operations) as a
\emph{`dirtier process'} (one who makes pages dirty).

    Linux maintains the page cache by marking the cached pages as active
    or inactive.
    A page that is accessed only once is an inactive page. Otherwise, the OS marks it
    as active. Linux cleans dirty pages by synchronizing least recently used
    inactive pages~\cite{activecache}.
    If the number of active pages grows too high, Linux marks some of the least
    recently used active pages as inactive.

    \subsection{Idle Instead of Direct Reclaim}

    Linux kernel versions older than version 3.10 applied
    the \emph{direct reclaim}~\cite{directreclaim} approach
    to prevent the \dirtyrate from exceeding the \dirtyratio. Whereby, as
    the \dirtyrate reaches the \dirtyratio, the \emph{dirtier process} is
    in charge of syncing the data to the storage device (foreground writeback).
    This approach keeps the \dirtyrate below the \dirtyratio.
    Though, in the case of
    IO to multiple inodes (open files), after reaching the
    \dirtyratio,
    many IOs to different places in the storage device (different
    inodes) will happen,
    requiring frequent seeking and resulting in significant performance
    degradation~\cite{fengguang}.

    Kernels starting from version 3.10 overcome this issue by throttling the
    dirtier process and letting it idle rather than making the process
    responsible for performing writebacks to the disk~\cite{fengguang,throttle}.
    Instead, the \emph{flusher threads} of the OS perform the writeback of the dirty data.
    Hence, the IO does jump more seldom to different inodes,
    and the OS manages the sync order of the dirty pages.

    \subsection{Throttling Regulation}
    \label {sec:throttle}

\begin{figure}[h!]
	\centering
	\includegraphics[width=.8\linewidth]{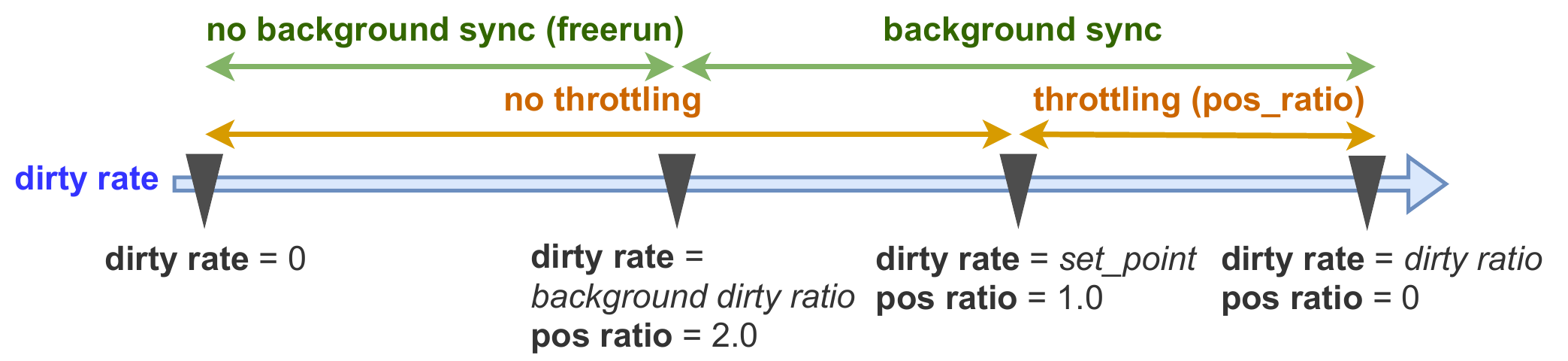}
    \caption{The meaning of different values of $\dirtyrate$ for the
      IO throttling of Linux kernels since version 3.10.\label{fig:io_time}}
    \end{figure}

    Another advantage of kernels starting from version 3.10 is the smooth
    IO slow down (throttling) of dirtier processes to maintain the \dirtyrate
    reasonably below the \dirtyratio.
    The kernel starts to throttle
    the dirtier processes when the \dirtyrate hits
    $\setpoint = (\bgrate + \dirtyratio) / 2$~\cite{setpoint}.
    As a result, the process observes
    a reduced rate of writing data (\emph{task rate} or bandwidth).
    In other words, the kernel limits the
    \emph{task rate} of a dirtier process by imposing a delay on it
    when it calls the $\func{write}$ system call.
    The kernel adjusts the \emph{task rate} of the process periodically as follows~\cite{posratio,throttle}:
    \begin{flalign}
        \label{eq:taskrate}
        \posratio &= 1 + {\big(\frac{\setpoint - \dirtyrate}{\dirtyratio - \setpoint}\big)}^3 \nonumber\\
        \emph{new task rate} &= \emph{average task rate} * \posratio
    \end{flalign}
    \Cref{eq:taskrate} tells how the kernel responds to a growing
    $\dirtyrate$ and how it
    limits the task rate of a dirtier process.
    The kernel keeps track of the average
    task rate for a dirtier process during a specific period and multiplies it by the
    computed $\posratio$ at that time to get the new task rate of the process.
    In the case of
    $\dirtyrate = \setpoint$, we have $\posratio = 1$ (start throttling), and in the case
    of $\dirtyrate$ hitting its maximum permitted value $\dirtyratio$, we get
    $\posratio = 0$ (no more writes).
    According to the new task rate, the kernel determines the amount of delay
    it should impose
    on the process before the $\func{write}$ system call returns, to
    meet the new computed task rate~\cite{rate_to_delay}.
    Notice that before the $\dirtyrate$ reaches
    $\setpoint$, no throttling occurs. Hence, $\posratio$ is not engaged to
    limit the task rate. \Cref{fig:io_time} represents an overview of
    the discussed throttling mechanism.

    \subsection{Storage Devices and Control Groups}

    When IO regards different storage
    devices, dirty pages destined to fast devices are
    synced more quickly than those destined to slow devices. As a result,
    the page cache will eventually be mostly filled with the dirty pages of
    slow devices syncing slowly while faster devices have a small share of dirty
    pages. To account for this, starting from the kernel version 2.6.24, each
    storage device gets a fair share of the
    $\dirtyratio$ according to its write-out rate (bandwidth)~\cite{dirtylimit_devices,dirtylimit_devices2}.
    Clearly, in the case of having most of the IO destined to a single storage device,
    that storage device gets nearly the entire share of the $\dirtyratio$.

    The kernels starting from version 4.10 also consider different control
    groups and penalize a control group for the dirty share of the inodes (files)
    owned by that control group~\cite{controlgroups}.
    A further improvement since kernel version 4.2
    penalizes the control group that is most responsible for the IO issued to an inode
    (not necessarily its owner): the ownership of such inode is
    changed to
    that control group~\cite{controlgroups2} (hence, penalizing the responsible control group).
    Again, in the simple case of having one control group or one process dirtying
    significantly more pages than the other control groups/processes, the
    throttling will be carried out mostly by that control group/process, and the
    other control groups can safely be ignored.

\section{Model} \label{sec:model}
\begin{figure}
        \tikzset{every picture/.style={line width=0.75pt}} 

\begin{tikzpicture}[x=0.75pt,y=0.75pt,yscale=-1,xscale=1]

\draw [line width=0.75]    (40.33,164.26) -- (82.18,164.26) ;
\draw [shift={(84.18,164.26)}, rotate = 180] [color={rgb, 255:red, 0; green, 0; blue, 0 }  ][line width=0.75]    (10.93,-3.29) .. controls (6.95,-1.4) and (3.31,-0.3) .. (0,0) .. controls (3.31,0.3) and (6.95,1.4) .. (10.93,3.29)   ;
\draw  [line width=0.75]  (95.57,151.77) -- (90.39,182.22) -- (85.21,151.77) -- (90.39,166.99) -- cycle ;
\draw  [draw opacity=0][line width=0.75]  (58.29,141.62) -- (68.3,141.62) -- (68.3,160.36) -- (58.29,160.36) -- cycle ; \draw  [line width=0.75]  (58.29,141.62) -- (58.29,160.36)(62.19,141.62) -- (62.19,160.36)(66.09,141.62) -- (66.09,160.36) ; \draw  [line width=0.75]  (58.29,141.62) --(68.3,141.62)(58.29,145.52) -- (68.3,145.52)(58.29,149.43) -- (68.3,149.43)(58.29,153.33) -- (68.3,153.33)(58.29,157.23) -- (68.3,157.23) ; \draw  [line width=0.75]   ;
\draw [line width=0.75]    (97.3,164.26) -- (139.14,164.26) ;
\draw [shift={(141.14,164.26)}, rotate = 180] [color={rgb, 255:red, 0; green, 0; blue, 0 }  ][line width=0.75]    (10.93,-3.29) .. controls (6.95,-1.4) and (3.31,-0.3) .. (0,0) .. controls (3.31,0.3) and (6.95,1.4) .. (10.93,3.29)   ;
\draw  [line width=0.75]  (152.53,151.77) -- (147.35,182.22) -- (142.18,151.77) -- (147.35,166.99) -- cycle ;
\draw  [draw opacity=0][line width=0.75]  (115.25,141.62) -- (125.26,141.62) -- (125.26,160.36) -- (115.25,160.36) -- cycle ; \draw  [line width=0.75]  (115.25,141.62) -- (115.25,160.36)(119.15,141.62) -- (119.15,160.36)(123.06,141.62) -- (123.06,160.36) ; \draw  [line width=0.75]  (115.25,141.62) -- (125.26,141.62)(115.25,145.52) -- (125.26,145.52)(115.25,149.43) -- (125.26,149.43)(115.25,153.33) -- (125.26,153.33)(115.25,157.23) -- (125.26,157.23) ; \draw  [line width=0.75]   ;
\draw [line width=0.75]    (153.22,165.04) -- (195.07,165.04) ;
\draw [shift={(197.07,165.04)}, rotate = 180] [color={rgb, 255:red, 0; green, 0; blue, 0 }  ][line width=0.75]    (10.93,-3.29) .. controls (6.95,-1.4) and (3.31,-0.3) .. (0,0) .. controls (3.31,0.3) and (6.95,1.4) .. (10.93,3.29)   ;
\draw  [line width=0.75]  (208.46,152.55) -- (203.28,183) -- (198.1,152.55) -- (203.28,167.77) -- cycle ;
\draw  [draw opacity=0][line width=0.75]  (171.17,142.4) -- (181.19,142.4) -- (181.19,161.14) -- (171.17,161.14) -- cycle ; \draw  [line width=0.75]  (171.17,142.4) -- (171.17,161.14)(175.08,142.4) -- (175.08,161.14)(178.98,142.4) -- (178.98,161.14) ; \draw  [line width=0.75]  (171.17,142.4) -- (181.19,142.4)(171.17,146.3) -- (181.19,146.3)(171.17,150.21) -- (181.19,150.21)(171.17,154.11) -- (181.19,154.11)(171.17,158.01) -- (181.19,158.01) ; \draw  [line width=0.75]   ;
\draw [line width=0.75]    (210.18,165.04) -- (252.03,165.04) ;
\draw [shift={(254.03,165.04)}, rotate = 180] [color={rgb, 255:red, 0; green, 0; blue, 0 }  ][line width=0.75]    (10.93,-3.29) .. controls (6.95,-1.4) and (3.31,-0.3) .. (0,0) .. controls (3.31,0.3) and (6.95,1.4) .. (10.93,3.29)   ;
\draw  [line width=0.75]  (265.42,152.55) -- (260.24,183) -- (255.06,152.55) -- (260.24,167.77) -- cycle ;
\draw  [draw opacity=0][line width=0.75]  (228.14,142.4) -- (238.15,142.4) -- (238.15,161.14) -- (228.14,161.14) -- cycle ; \draw  [line width=0.75]  (228.14,142.4) -- (228.14,161.14)(232.04,142.4) -- (232.04,161.14)(235.94,142.4) -- (235.94,161.14) ; \draw  [line width=0.75]  (228.14,142.4) -- (238.15,142.4)(228.14,146.3) -- (238.15,146.3)(228.14,150.21) -- (238.15,150.21)(228.14,154.11) -- (238.15,154.11)(228.14,158.01) -- (238.15,158.01) ; \draw  [line width=0.75]   ;
\draw [line width=0.75]    (270.45,166.04) -- (312.3,166.04) ;
\draw [shift={(314.3,166.04)}, rotate = 180] [color={rgb, 255:red, 0; green, 0; blue, 0 }  ][line width=0.75]    (10.93,-3.29) .. controls (6.95,-1.4) and (3.31,-0.3) .. (0,0) .. controls (3.31,0.3) and (6.95,1.4) .. (10.93,3.29)   ;

\draw (68.9,60.42) node [anchor=north west][inner sep=0.75pt]  [font=\small,rotate=-90] [align=left] {prepare data 0};
\draw (96.72,82.62) node [anchor=north west][inner sep=0.75pt]  [font=\small,rotate=-90] [align=left] {write to disk};
\draw (127.05,61.42) node [anchor=north west][inner sep=0.75pt]  [font=\small,rotate=-90] [align=left] {prepare data 1};
\draw (153.68,82.62) node [anchor=north west][inner sep=0.75pt]  [font=\small,rotate=-90] [align=left] {write to disk};
\draw (181.79,61.2) node [anchor=north west][inner sep=0.75pt]  [font=\small,rotate=-90] [align=left] {prepare data 2};
\draw (209.61,83.4) node [anchor=north west][inner sep=0.75pt]  [font=\small,rotate=-90] [align=left] {write to disk};
\draw (238.75,62.2) node [anchor=north west][inner sep=0.75pt]  [font=\small,rotate=-90] [align=left] {prepare data 3};
\draw (266.57,83.4) node [anchor=north west][inner sep=0.75pt]  [font=\small,rotate=-90] [align=left] {write to disk};
\draw (54.34,170.4) node [anchor=north west][inner sep=0.75pt]    {$d_{0}$};
\draw (68.79,188) node [anchor=north west][inner sep=0.75pt]    {$o_{0} :s_{0}$};
\draw (110.43,170.4) node [anchor=north west][inner sep=0.75pt]    {$d_{1}$};
\draw (126.48,188.8) node [anchor=north west][inner sep=0.75pt]    {$o_{1} :s_{1}$};
\draw (166.53,171.4) node [anchor=north west][inner sep=0.75pt]    {$d_{2}$};
\draw (182.37,188) node [anchor=north west][inner sep=0.75pt]    {$o_{2} :s_{2}$};
\draw (223.21,171.4) node [anchor=north west][inner sep=0.75pt]    {$d_{3}$};
\draw (239.07,188.8) node [anchor=north west][inner sep=0.75pt]    {$o_{3} :s_{3}$};
\draw (279.9,171.4) node [anchor=north west][inner sep=0.75pt]    {$d_{4}$};
\draw (294.59,60.5) node [anchor=north west][inner sep=0.75pt]  [font=\small,rotate=-90] [align=left] {delay before close};
\draw (335.67,163.75) node  [font=\small] [align=left] {\begin{minipage}[lt]{30.43pt}\setlength\topsep{0pt}
file close
\end{minipage}};
\draw (35.91,163.75) node  [font=\small] [align=left] {\begin{minipage}[lt]{30.43pt}\setlength\topsep{0pt}
file open
\end{minipage}};

\end{tikzpicture}
 
        \caption{A process opens a file, writes data in $4$ chunks to the file
            at the offsets $o_0, o_1, o_2, o_3$ with
            sizes $s_0, s_1, s_2, s_3$ (after computing each chunk), and
            closes the file\label{fig:processmodel}.}
    \end{figure}
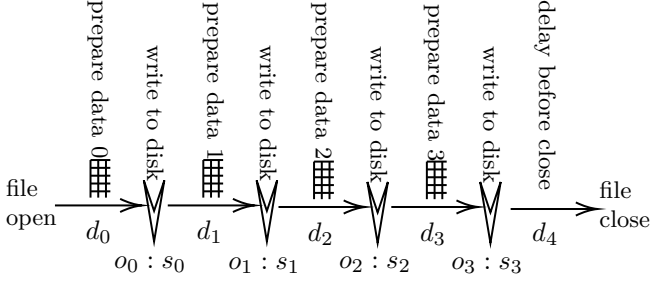

    In this work, we provide a model to predict the expected write
    throughput of a process. We assume a single IO-intensive process
    issues \emph{write} requests to a single storage device. Since
    there are no other IO-intensive processes, nearly all IO requests come from
    the control group that owns the IO-intensive process. As a result, the global
    $\posratio$ determines the task rate of the process.
    Later, we will discuss possibilities to expand our model
    for more complicated scenarios.
    As discussed, Linux performs background syncs when the \emph{dirty rate}
    reaches the $\bgrate$, and when it exceeds the $\dirtyratio$, Linux throttles
    the dirtier process,
    as discussed in \cref{sec:throttle}.
    The process could engage library calls or
    system calls to issue the writes. The IO could follow a random or a sequential
    pattern.
    We will also model the effects of opening a file with
    the \ODIRECT and \OSYNC flags \cite[open]{man2}.
    Our model estimates the IO costs of a dirtier process that opens a
    file and performs $n$ write operations in the file with different offsets and sizes.
    Additionally, the process requires different computation
    times $\delay\ge0$ before performing each write operation.
    \Cref{fig:processmodel} illustrates such a process.
    Throughout this work, we use the term \emph{cost} of an
    operation to refer to the \emph{duration} of performing that operation.
    We denote the total data size that the process writes as $S$.

    \subsection{Basic System Parameters} \label{sec:basic-parameters}
    On the system side, the main memory has a bandwidth of
    $\mbw$, the Ramdisk accepts data with a rate of $\rambw$,
    and the storage device has a sync bandwidth of
    $\sbw$ and a read bandwidth of $\srbw$.
    The filesystem's logical block size on the underlying storage
    device is $\bs$. The C standard library has a
    buffer size of $\cbf$ in the user space to buffer the data passed to $\func{fwrite}$.
    The operating system starts
    background flushes when the $\dirtyrate$ reaches $\bgrate$, starts throttling
    when reaching $\setpoint=(\bgrate+\dirtyratio)/2$, and has a hard limit
    of $\dirtyratio$, as discussed in \cref{sec:throttle}. Moreover, the OS
    cleans dirty pages older than $\dirtyexpire$ seconds in the background.
    Notice that we differentiate $\mbw$ and $\rambw$, even though
    they both write to the same device (main memory). The reason is that
    to write data on the Ramdisk, the OS comes in between and performs many
    other operations and updates the file metadata for each written
    page~\cite[Sect.~12--13]{love2010linux_devel},\cite{stackex}.
    For instance, the kernel allocates memory for written
    data (memory allocation can be costly). It also updates the inode data
    structures to update the file size and keep track of each inode's
    allocation. Hence, a Ramdisk has
    a lower bandwidth than plain memory access ($\rambw < \mbw$).
    We denote the latency of invoking a $\func{write}$ system call as
    $\syswcost$, and that of a
    \emph{sync} $\func{write}$ system call (when
    synced data access is specified, e.g., by \OSYNC)
    as $\sysswcost$ (the extra fixed cost of the non-sync
    and sync write system calls).
    The average cost of seeking the storage device
    (moving the disk head around) is represented by $\skcost$.
    To perform our evaluations, we implemented a micro-benchmark in
    \Cpp that automatically measures the values of these basic system
    parameters on a given platform. In \cref{sec:eval} we explain the
    approach of our benchmarking tool.

    \subsection{Modeling Synchronized and Direct IO with System Calls}
    \label{sec:mod-sync-direct-syscall}
\begin{figure*}
        \centering
        \scalebox{.8}{\input{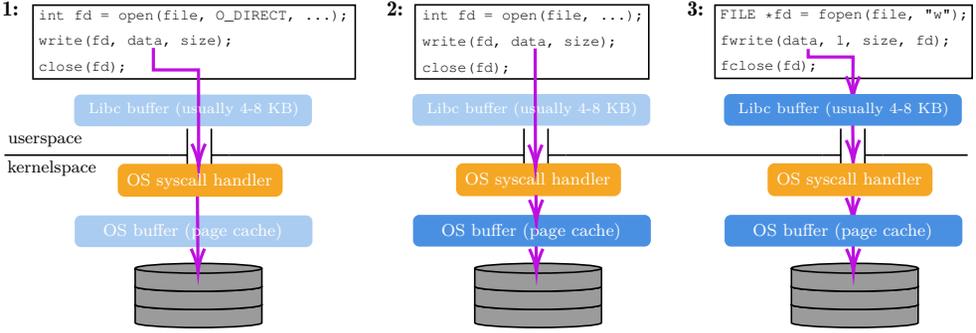}}
        \caption{\textbf{1}: Direct IO bypassing the user space and OS
            cache and going straight to the storage device.
            \textbf{2}: IO issued by system-calls bypassing the
            user space and passing through the OS page cache.
        \textbf{3}: IO issued by library functions passing through the library
        buffer in the user space and the OS buffer (page cache) before going
        to the storage device.
        \label{fig:iopath}}
    \end{figure*}

    If a file open uses the \ODIRECT flag, the IO for that file bypasses the OS
    page cache and goes directly to the storage device without being cached
    in user or kernel space, as illustrated in case 1 of~\cref{fig:iopath}.
    The \ODIRECT flag guarantees that the OS does not copy the data
    into the kernel space.
    Although, it does not promise that write operations return strictly after
    transferring all the data to the storage device.
    Such a guarantee is made by the \OSYNC and \ODSYNC flags, while
    those do not
    promise to bypass the OS cache~\cite[open]{man2}. Combining
    \ODIRECT with either of these flags guarantees a
    \emph{synchronized} and \emph{direct} write to the storage device.
    Another way to achieve synchronized \emph{and} direct IO is to
    open the file with the
    \ODIRECT flag and calling the $\func{fsync}$ system call
    after each $\func{write}$ system call to make sure that
    the data is synchronized to the storage device. Since \ODIRECT bypasses the OS and sends the data directly to the
    storage device, a limitation is that the data size must be a multiple of
    the logical block size of the filesystem~\cite[open]{man2}.

    Estimating the IO cost (IO duration) for synchronized and
    direct IO is simpler than with
   caching levels involved, as we will see later.
    As the data goes directly to the storage device, the only overhead
    is the system-call handler of the OS.
    As a result, the cost of writing a single data chunk
    comes from two
    sources: first, the cost of making the sync $\func{write}$ system call
    $\sysswcost$, and second, the cost of writing the data to the
    storage device. Additionally, if the data access follows a
    random pattern ($\israndom = 1$, otherwise, $\israndom = 0$),
    we have extra seeking cost $\skcost$, as well.
    Hence, the IO cost of writing data of size $S$ in $n$ chunks
    (chunks may have different sizes) is defined as follows:
    \begin{flalign}
        T^{\mathit{direct}} &=
        \sysswcost n + \israndom\cdot \skcost n+ \frac{S}{\sbw}
        \label{eq:directcost}
    \end{flalign}

    \Cref{eq:directcost} estimates the total IO cost when writing
    data of size $S$ in $n$ chunks using the direct and synchronized flags.
    As $n$ grows (data is written in smaller pieces), the observed
    cost grows because of the growing cost of system calls and seeks
    (when doing random access). Later, in \cref{sec:eval}, we compare
    this model with practical evaluation experiments.

    \subsection{Modeling Synchronized and Non-direct IO with System Calls}
    \label{sec:mod-sync-nondirect-syscall}

    In the absence of the \ODIRECT flag, the OS copies the data
    into the page cache before writing it to the storage device.
    But still, when \OSYNC or \ODSYNC is passed to the $\func{open}$ system call,
    the $\func{write}$ call guarantees to return only when the data is synced
    to the storage device (after passing through the page cache).
    Alternatively, the process can open the file without these flags but enforce synchronization by calling the
    $\func{fsync}$ system call after each $\func{write}$.
    The second case of \cref{fig:iopath} shows the path of data in this scenario.

    The extra copying of data from the user space (application) into the
    kernel space (page cache) imposes a minor additional cost on the IO, compared
    to the direct and synchronized IO. As the IO is non-direct and
    passes through the OS, data sizes do not have to be multiples of the block size.
    The $\func{write}$ system call now accepts any data size $s_i\in\{s_0, s_1, ..., s_{n-1}\}$ ,
    where $\sum_{i=0}^{n-1}s_i = S$.
    The OS manages the data access and adjusts the data size, as follows:
    If $s_i$ is a multiple of the logical block size of the
    filesystem $\bs$ ($\isfit_i=1$),
    the OS transfers the data to the storage device.
    Otherwise ($\isfit_i=0$), first, the OS transfers the data in chunks of size
    $\bs$ as much as possible. For the remaining data $\just{rem_i} = s_i\cmod\bs$
    (smaller than block size),
    the OS reads the corresponding block from the storage
    device (with the cost of $\bs/\srbw$),
    applies the remaining data on the read block ($\just{rem_i}/\rambw$),
    and writes the updated block back to the storage device
    ($\bs/\sbw$)~\cite{blocksize_penalty}.
    Notice the extra cost penalty for
    the non-fitting data sizes. The rest remains as in the case of
    synchronized and direct IO:
    \begin{flalign}
        \just{rem_i} &= s_i\cmod\bs\nonumber\,, \hspace{50pt}
        \just{fit_i} = s_i - \just{rem_i}\nonumber\,,\\
        \syncti &= \sysswcost+\israndom\cdot\skcost + s_i/\rambw +
            \frac{\just{fit_i}}{\sbw}\nonumber\\
            &+ \big[\frac{\bs}{\srbw} + \frac{\bs}{\sbw}\big]\cdot\isfit_i
        \label{eq:sync_chunkcost}
    \end{flalign}

    \Cref{eq:sync_chunkcost} represents the observed cost when performing a synchronized
    non-direct IO of size $s_i$, where $\just{fit_i}$ is the data part that is
    transferred to the storage device in $\bs$ chunks, and $\just{rem_i}$ is the
    remaining non-fitting part of the data for which we have a cost penalty.
    If the data size is divisible to the blocksize $\bs$, then $\isfit_i = 0$,
    and the equation is still valid. Here, we also see the extra cost of
    copying the data into the page cache before
    transferring it to the storage device. Following \cref{eq:sync_chunkcost},
    the IO cost of writing the whole data consisting of $n$ chunks
    with the sizes $s_0, s_1, ..., s_{n-1}$ is defined as:
    \begin{flalign}
        T^{\mathit{sync}} &= \sum_{i=0}^{n-1} \syncti
        = \sysswcost n + \israndom\cdot \skcost n+ \frac{S}{\sbw} \\
                 &+ \hspace{-8pt}\underbrace{\frac{S}{\rambw}}_{\text{copy to kernel space}}
        \hspace{-8pt}+ \underbrace{\sum_{i=0}^{n-1} \big[\frac{\bs}{\srbw} +
        \frac{\bs}{\sbw}\big]\cdot\isfit_i -
         \frac{\just{rem_0}}{\sbw}}_{\text{non-fitting size penalty}}\nonumber
        \label{eq:synccost}
    \end{flalign}

    Later, in \cref{sec:eval}, we will evaluate these equations with
    our IO experiments and demonstrate the extra cost of non-fitting
    data sizes.

    \subsection{Modeling Non-synchronized IO with System Calls}
    \label{sec:mod-nonsync}

    If none of the \ODIRECT, \OSYNC, or \ODSYNC flags is specified,
    the OS copies the data to the page cache (see
    the second case of \cref{fig:iopath}) as \emph{dirty pages}
    instead of immediately transferring them
    to the storage device. Then, it is the responsibility of the OS
    to sync the data to the storage device in the background and keep the
    \dirtyrate between the \bgrate and \dirtyratio by adequately throttling
    the dirtier process, as discussed in \cref{sec:throttle}.
    Before throttling begins, the process effectively observes
    $\rambw$, as the OS merely writes the data on the page cache and
    updates the data structures representing the target file.

    Performing background synchronization is an expensive task, as the
    OS has to go through the pages and find appropriate candidate
    pages to be cleaned. The kernel is busier than usual, consumes CPU
    power, and accesses the main memory in the background.  As a
    result, the process generating dirty pages in the page
    cache (writer) observes a reduced IO throughput to a certain extent when
    the kernel performs background synchronization (e.g., when the
    $\dirtyrate$ exceeds the $\bgrate$).  This is just a side-effect, as
    the kernel did not explicitly decide to throttle the process
    (e.g., when $\dirtyrate < \setpoint$). We consider this effect
    and denote the reduced bandwidth of ramdisk under the OS background
    synchronization as $\redbw$.
    \Cref{sec:eval} will show this effect on different hosts and for
    different scenarios.
    Also, we implemented a tool to measure $\redbw$ of a
    platform.

    In contrast to the previous cases, delays before
    IOs become relevant for non-synchronized IO. Here, the OS
    cleans dirty pages in the background during those delays
    and determines the \emph{task rate}
    (IO throughput) according to the number of dirty pages on
    the page cache (see \cref{sec:throttle}). The OS determines the values
    of $\bglimit$ and
    $\hlimit$ by multiplying the number of dirtyable pages by the
    $\bgrate$ and $\dirtyratio$, as discussed in \cref{sec:throttle}.
    Though, $\bglimit$ and $\hlimit$ can also directly be fetched from
    the file \emph{/proc/vmstat}~\cite{vmstat}
    under Linux without computing the number of dirtyable pages
    and multiplying it with the $\bgrate$ and $\dirtyratio$.
    The OS cleans the dirty pages older than the $\dirtyexpire$. The value of
    $\dirtyexpire$ is fetched either using the \emph{sysctl} command
    (under the name \emph{dirty\_expire\_centisecs}) or
    by reading the file \emph{/proc/sys/vm/dirty\_expire\_centisecs} under
    Linux~\cite{dirtylimits}.
    If the number of $\dirty$ pages is below $\bglimit$,
    and there are no expired pages in the page cache, the
    OS does not perform background
    synchronization. We denote this state as $\freerun$. Otherwise, if
    the number of $\dirty$ pages exceeds $\bglimit$ but is still below
    $\setpoint$, or if there are expired pages, then the OS performs
    background flushes. We call this state $\asyncrun$. Lastly, if
    the number of $\dirty$ pages exceeds $\setpoint$, not only does the OS
    perform background flushes, but it also throttles the dirtier process by
    reducing its \emph{task rate} (IO throughput).
    As a result, the process observes a higher IO cost.
    This state is denoted as $\throttlerun$.

    In the following, we estimate the cost of non-synchronized IO with
    system calls. The behavior of IO throttling is too complex to fit in a
    single equation as in the previous cases. Therefore, we provide an algorithm
    to calculate the cost of non-synchronized IO in the $\nextsysbw$ procedure.

    \paragraph{Procedure \nextsysbw{} (see \cref{alg:nonsync_bw})}

    The procedure gets the $\file$, $\offset$, and the $\size$ of the
    data chunk, as well as the $\delay$ before writing it,
    as inputs and estimates the observed cost of writing the data chunk.
    To approximate the IO cost of the $i$th data chunk, $\just{\nextsysbw}$
    should be called first for all the data chunks that happened before it, i.e.,
    the data chunks $\{0, 1, ..., i-1\}$. The reason is that, by each invocation,
    $\just{\nextsysbw}$ updates the global variables, such as the number
    of dirty pages and the average bandwidth. These global variables regulate
    the upcoming estimations.
    If the $\file$, $\offset$s, $\sizes$ and $\delays$ are known in advance, looping through
    them and calling $\just{\nextsysbw(\file, \offset, \sizes, \delay)}$ predicts the costs for all chunks.
    Otherwise, $\just{\nextsysbw(\file, \offset, \size,\delay)}$ can be called as soon as the
    $\offset$ and $\size$ of the next data chunk of a $\file$ and its $\delay$ are known.

    The procedure updates the $\ttime$ considering
    the $\delay$ and performs
    a background flush during the $\delay$
    if the state is not $\freerun$ (in
    $\freerun$, $\bgflush{}$ returns immediately).
    In \cref{line:exp} of \cref{alg:nonsync_bw},
    $\haveexpired$ indicates the existence of expired pages.
    In the case of $\dirtyrate <\setpoint$ (no OS throttling), if the \dirtyrate
    exceeds the \bgrate or
    some pages are expired
    (\cref{line:async}: $\asyncrun$), the process observes
    $\redbw$, as the OS is performing background sync (but does not
    explicitly throttle the process).
    Otherwise, if $\dirtyrate \ge\setpoint$ (\cref{line:throttle}:
    $\throttlerun$), the OS throttles the IO throughput and determines the
    \emph{task rate} of the process by multiplying the average bandwidth
    $\avgbw$ (average task rate) by $\posratio$,
    as discussed in \cref{sec:throttle} (see \cref{line:posratio}).
    If the state is neither of the $\asyncrun$ nor $\throttlerun$,
    we are in $\freerun$,
    and the \emph{task rate} stays $\rambw$. Because, during $\freerun$, the
    $\func{write}$ system call returns as soon as the data chunk is
    copied into the page cache, the process effectively observes the
    Ramdisk bandwidth $\rambw$.

\begin{figure}[htb]
  \centering
  \begin{minipage}{.67\linewidth}
    \begin{algorithm}[H]
\begin{spacing}{1}
    \small
	\caption{Non-sync Syscall IO Cost Estimation}
	\label{alg:nonsync_bw}
\begin{flushleft}
    \Data{$\sbw$ \& $\rambw$: storage device and Ramdisk bandwidth}
    \Datan{$\bglimit$ \& $\hlimit$: dirty background \& dirty thresholds}
    \Datan{$\dirtyexpire$: dirty page expire time}
    \Datan{$\redbw$: ramdisk bandwidth under OS background sync}
    \hspace*{\algorithmicindent}\textbf{Struct: }\just{\datablock\,\,\{\file, \offset, \size, \ioendtime, \activedata\}}\\
    \Global{$\iotime \gets 0$, $\ttime \gets 0$, $\pagecache\gets$ \textbf{empty} \datablock\textbf{ list}}
    \Globaln{$\dirty \gets 0$, $\avgbw \gets 0$, $\setpoint \gets(\bglimit  + \hlimit)/2$}
    \Param{$\file, \offset, \size$: the metadata of the data chunk}
    \Paramn{$\delay$: the delay before writing the data chunk}
  \end{flushleft}
    \begin{algorithmic}[1]
        \Procedure{\nextsysbw}{}
            \State $\ttime \pluseq \delay$,
                \hfill$\bw \gets \rambw$
            \State $\bgflush{\delay, \ttime}$
            \State $\haveexpired \gets$ \expired\label{line:exp}
            \If{$\dirty < \setpoint\,\land(\dirty \ge \bglimit \lor\, \haveexpired)$}\label{line:async}\comdoc{async}
                \State $\bw \gets \redbw$
            \ElsIf{$\dirty \ge \setpoint$}\label{line:throttle}\comdoc{throttle}
                \State $\posratio \gets 1 +
                    {\big(\frac{\setpoint - \dirty}
                    {\hlimit - \setpoint}\big)}^3$\label{line:posratio}
                \State $\bw \gets \just{min}(\avgbw \cdot \posratio, \redbw)$
            \EndIf\label{line:bw2}
            \State $\iocost \gets \size/\bw + \syswcost$
            \State $\updatecache{\textbf{new }\datablock\langle\file, \offset, \size\rangle}$\label{line:upcache}
            \State $\avgbw \gets (\avgbw\cdot\iotime + \size) \big/ (\iotime + \iocost)$ \label{line:avgbw}
            \State $\ttime \pluseq \iocost$,\hfill$\iotime\pluseq\iocost$\label{line:uptime}
            \State $\bgflush{\iocost, \ttime}$
            \State \Return $\iocost$
        \EndProcedure
        \Procedure{$\updatecache{\newdblock}$}{}
            \State $\matchingblocks \gets \overlapfn{\pagecache, \newdblock}$\label{line:overlap}
            \State $\dirty \pluseq \newdblock.\size$\hfill$\offset\gets\newdblock.\offset$
            \ForAll{$\block \in \matchingblocks$}\label{line:cachefor1}

                \State $\nextlim \gets (\just{b_{i+1}} \textbf{ is null}) \textbf{ ? } \just{inf} \textbf{ : } \just{b_{i+1}}.\offset$
                \State $\limblock \gets \cutblock{\newdblock,\offset, \nextlim}$
                \State $\overlap \gets \block\cap\newdblock$,\hfill$\overlap.\activedata\gets$\textbf{true}
                \State $\cachedisj\gets \block - \overlap$
                \State $\newdisj\gets \limblock - \overlap$
                \State $\forall d \in \cachedisj;d.\activedata \gets \block.\activedata,\hfill\pagecache.\insertlist{d}$
                \State $\forall d \in \newdisj;\,d.\activedata\gets\textbf{false},\hfill\pagecache.\insertlist{d}$
                \State $\pagecache.\insertlist{\overlap}$,\hfill$\pagecache.\just{remove}(\block)$
                \State $\dirty \minuseq \overlap.\size$,\hfill$\offset\gets\nextlim$
            \EndFor\label{line:cachefor2}
        \EndProcedure
        \algstore{algbreak}
      \end{algorithmic}
      \hfill (continued $\to$)
\end{spacing}
    \end{algorithm}
  \end{minipage}
\end{figure}

\begin{figure}[htb]
  \centering
  \begin{minipage}{.67\linewidth}
    \begin{algorithm}[H]
	\begin{spacing}{1}
		\small
	\caption*{\textbf{Algorithm 1} Non-sync Syscall IO Cost Estimation (continued)}
  	\begin{algorithmic}[1]
        \algrestore{algbreak}
        \Procedure{$\bgflush{\period, \ttime}$}{}
  		\While{$(\lnot\pagecache\just{.empty()})\land(\dirty \ge \bglimit\,\lor$
  			\Statex\hspace{\algorithmicindent}\hspace{10pt}
  			\expired)}\label{line:bgwhile1}
  		\State $\balancecache{}$\comdoc{make some old active data inactive}
  		\State $\tobecleaned \gets $\emph{ the oldest inactive data block in \pagecache}\label{line:evictable}
  		\If{$\period \ge \tobecleaned.\size / \sbw$}\label{line:ifevictall}
  		\State $\period \minuseq \tobecleaned.\size / \sbw$
  		\State $\dirty \minuseq \tobecleaned.\size$,\hfill
  		$\pagecache.\just{remove}(\tobecleaned)$
  		\Else
  		\State $\tobecleaned.\size \minuseq
  		\period\cdot\sbw$\label{line:update_size}
  		\State $\dirty \minuseq \period\cdot\sbw$
  		\State $\textbf{break}$
  		\EndIf
  		\EndWhile\label{line:bgwhile2}
  		\EndProcedure
        \algstore{algbreak2}
    \end{algorithmic}
\end{spacing}
    \end{algorithm}
  \end{minipage}
\end{figure}

    Next, \cref{line:upcache} inserts the metadata of the
    data chunk into the page cache model (explained in the next section).
    \Cref{line:avgbw} maintains the average bandwidth $\avgbw$ according
    to~\cite{avgbw}, and \cref{line:uptime} updates $\ttime$ and $\iotime$
    (time of performing IO
    without the delays). Lastly, $\bgflush{}$ is called again
    to take account of the pages that were cleaned
    during $\iocost$ in the background
    and to update the number of dirty pages accordingly.
    If the state is $\freerun$,
    $\bgflush{}$ does nothing and returns immediately.

    \subsubsection{$\updatecache{}$}
    We model the
    granularity of data in the page cache as $\datablock$s, i.e.,
    data structures of $\{\file, \offset, \size, \ioendtime, \activedata\}$.
    $\file$, $\offset$, and $\size$ describe the location of data, $\ioendtime$
    indicates the finish time of the IO, and $\activedata$ specifies if the data
    is in the active or inactive cache.
    Our cache model keeps the metadata as $\datablock$s rather than
    keeping track of every page. This is similar to the approach of
    Hoang-Dung~\cite{do2021modeling} that extends the WRENCH workflow
    simulator~\cite{wrench} to
    take the Linux cache into account.
    To insert the metadata of the new $\datablock$ ($\newdblock$)
    in the cache, the $\updatecache{}$ procedure finds the blocks
    in the cache that overlap with the $\newdblock$,
    considering the
    $\file$, $\offset$, and $\size$
    (\cref{line:overlap}).
    The $\just{for}$ loop at \crefrange{line:cachefor1}{line:cachefor2}
    loops through $\block \in \matchingblocks$ (overlapping blocks),
    and for each pair of $\block$ and $\newdblock$, splits the data
    into the overlapping and
    non-overlapping (disjunct) parts. Different parts of the data are again inserted
    in the active or inactive cache as follows:
    The non-overlapping parts of $\block$ with unchanged $\activedata$,
    the non-overlapping parts of $\newdblock$ with $\activedata = \textbf{false}$
    (inactive cache), as they are accessed only once (first access), and
    the overlapping part with $\activedata = \textbf{true}$ (active cache),
    as they are accessed more than once.

    \paragraph{$\bgflush{}$} The parameter
    $\period$ indicates the duration we have for background flushing.
    The $\just{while}$ loop at \crefrange{line:bgwhile1}{line:bgwhile2}
    repeats as long as
    $\dirty$ pages exceed $\bglimit$ or there are expired pages (as long as
    sync is to be performed).
    The $\balancecache{}$ procedure balances the number of active and
    inactive pages by marking some old active
    pages as inactive~\cite{activecache}.
    \Cref{line:evictable} chooses an inactive $\datablock$ with the
    lowest $\ioendtime$ (the oldest inactive data) as the next $\datablock$ in
    the cache to be evicted ($\tobecleaned$).
    If during the $\period$ $\tobecleaned$ can be synced completely
    (\cref{line:ifevictall}), it is removed from
    the cache, otherwise, $\tobecleaned.\size$ is reduced (synced) as much
    as the $\period$ allows (\cref{line:update_size}).
    Notice that the seeking cost and the block size penalty are
    absent when estimating the cost of non-synchronized IO.
    The page cache accumulates the IO operations smaller than
    the block size with other accesses to the same block and transfers
    them to the storage device in the granularity of pages.
    Also, even when the OS has to perform seeks or faces
    block size penalties, it
    all happens in the background and is transparent to the process.

    \subsection{Modeling Standard-Library IO}\label{sec:mod-libcost}

    The C standard library provides a set of IO functions
    on top of the system calls, as discussed in \cref{sec:libcall}. The library
    IO functions work with a buffer of size $\cbf$ in the user space.
    $\func{fwrite}$ writes the data into the user space buffer instead of immediately
    invoking a system call to perform the IO.
    This way, it spares the cost of making frequent system calls.
    When the buffer is full, $\func{fwrite}$ invokes a
    system call (e.g., $\func{write}$) to transfer the buffer's aggregated data
    into the page cache and empty the buffer.
    The third case of \cref{fig:iopath}
    illustrates the data path to the storage device in this scenario. Same as
    the model of non-synchronized IO with system calls, estimating the IO cost for
    library IO is too complex to fit in a single equation. Hence, we provide the
    $\just{\nextlibbw}$ procedure to calculate the IO cost for library IO.

    The procedure $\just{\nextlibbw}$ (see \cref{alg:lib_bw}) gets
    the $\file$, $\offset$, and the $\size$ of the
    data chunk, as well as the $\delay$ before writing it,
    as inputs and estimates the observed cost of writing the data chunk.
    Similar to $\just{\nextsysbw}$, the data chunks should be passed to
    $\just{\nextlibbw}$ in order.

    First, at \cref{line:checkseq}, the procedure checks whether the
    IO is sequential (i.e., $\func{fseek}$ is
    not called and the $\offset$ points to the subsequent location of the
    last IO, $\prevpos$). If the $\offset$ is changed (random access), the
    buffer must be flushed to the OS page cache before filling it with the new
    data. Accordingly, the model
    accounts for the cost of invoking the system call at \cref{line:seekflush}.
    Here, $\prevoffset$ points
    to the offset of buffered data, $\pending$ indicates
    the data size in the buffer, and $\pendingd$ keeps track of the
    accumulated delay until the next system call invocation
    ($\just{\nextsysbw}$).
    Next, if the data fits in the buffer (\cref{line:bufferfit}), the
    process merely observes
    the cost of writing the data in the user space buffer ($\size / \mbw$).
    Accordingly, $\just{\nextlibbw}$ updates the data size in the buffer
    (the global variable $\pending$) and accumulates the time
    (delay) until the next system
    call (the global variable $\pendingd$).

\begin{figure}[htb]
  \centering
  \begin{minipage}{.67\linewidth}
    \begin{algorithm}[H]
\begin{spacing}{1}
    \small
	\caption{Standard Library IO Cost Estimation}
	\label{alg:lib_bw}
\begin{flushleft}
    \Data{$\mbw$: memory bandwidth}
    \Datan{$\cbf$: size of the standard library IO buffer}
    \Global{$\pending\gets 0$, $\pendingd\gets0$, $\prevpos \gets 0$}
    \Param{$\file$, $\offset$, $\size$: the metadata of the data chunk}
    \Paramn{$\delay$: the delay before writing the data chunk}
\end{flushleft}
    \begin{algorithmic}[1]
        \algrestore{algbreak2}
        \Procedure{\nextlibbw}{}
        \If{$\offset \ne \prevpos$}\label{line:checkseq}
            \State $\prevoffset \gets \prevpos - \pending$
            \State $\iocost \gets \nextsysbw\just{(\file, \prevoffset, \pending, \pendingd)}$\label{line:seekflush}
            \State $\pending \gets 0$
        \EndIf
            \If{$\size \le \cbf - \pending$}\label{line:bufferfit}
                \State $\iocost \pluseq \size/\mbw$,
                \hfill$\pending \pluseq \size$
                \State $\pendingd \pluseq \delay + \iocost$

            \Else
                \State $\iocost \pluseq (\cbf - \pending)/\mbw$\label{line:restbuffer}
                \State $\pendingd \pluseq \delay + \iocost$,
                \hfill$\prevoffset \gets \offset - \pending$
                \State $\iocost\pluseq \nextsysbw\just{(\file, \prevoffset, \cbf, \pendingd)}$\label{line:sysbuf}
                \State $\rest \gets \size - \cbf + \pending$,
                \hfill$\rem \gets \rest \cmod \cbf$
                \If{$\rest \ge \cbf$}\label{line:stilllarger}
                    \State $\iocost \pluseq \nextsysbw\just{(\file, \prevoffset + \cbf, \rest - \rem, 0)}$\label{line:sysrest}
                \EndIf
                \State $\iocost \pluseq \rem/\mbw$,
                \hfill$\pending \gets \rem$\label{line:leftover}
                \State $\pendingd \gets \rem/\mbw$

            \EndIf
            \State $\prevpos \gets \offset + \size$
            \State \Return $\iocost$
        \EndProcedure
    \end{algorithmic}
\end{spacing}
   \end{algorithm}
  \end{minipage}
\end{figure}

    Otherwise, if the data does not fit in the buffer, $\func{fwrite}$ fills
    the buffer with the data as far as
    possible (modeled at \cref{line:restbuffer}) and invokes a system call
    to write the buffer in the page cache, as modeled at \cref{line:sysbuf}.
    Next, if the leftover data ($\rest$) is still larger than the
    buffer (\cref{line:stilllarger}), $\func{fwrite}$ invokes a
    second system call to write the remaining
    data (adjusted according to $\cbf$) at once to the page cache
    without buffering (modeled at \cref{line:sysrest}). Notice that $\func{fwrite}$
    invokes the second system call after the first one, at \cref{line:sysbuf},
    without delay ($\delay=0$ passed to $\nextsysbw$). Finally, if
    the $\rest$ size was not a multiple of $\cbf$, $\func{fwrite}$
    buffers the leftover (modeled at \cref{line:leftover}). Notice that throughout
    all these steps, our algorithm accumulates $\pendingd$ accordingly.
    If $\func{fwrite}$ is called to write large chunks, the observed
    cost converges to the cost of the $\func{write}$ system call
    because in that case $\func{fwrite}$ writes most of the data
    without buffering them in the user space (see \cref{line:sysrest}).
    The minor extra cost of copying several kilobytes of the
    leftovers in the user space buffer with the plain memory bandwidth
    $\mbw$ is negligible compared to the cost of a system call that writes large data.
    By calling $\func{fflush}$, a system call is invoked to
    move the buffer data to the page cache and empty the buffer.

    Notice that the IO operations of other libraries or languages
    (e.g., $\func{fstream}$ of \Cpp) buffer the data up to several
    kilobytes, as well, and follow almost similar concepts as $\just{\nextlibbw}$.
    Hence, our algorithm is not limited to the IO operations of
    the C standard library but can also be adjusted to estimate the IO
    cost in other cases with fixed buffer sizes.

\section{Related Work} \label{sec:rel}

SimGrid~\cite{simgrid} is a simulation framework that models the performance of distributed
and undistributed applications. It can model the CPU, network, IO, etc. and
estimate the cost of running applicatrions on a specific host. To model
the performance of IO operations, SimGrid disregards the OS caches and divides the
data size to the bandwidth of the storage device (na\"ive approach). In \cref{sec:eval},
we compare our model CAWL with SimGrid regarding the IO modeling and show the extent of
inaccuracy of the SimGrid model and the importance of considering cache effects.

Hoang-Dung et al.~\cite{do2021modeling} propose an IO simulation model and consider
some properties of the Linux page cache mechanism. They aim to improve the
accuracy of the WRENCH workflow simulation framework~\cite{wrench} by taking the
Linux page cache into account. They simulate the
behavior of the active and inactive cache of Linux to keep track
of data blocks, whether they are read from the page cache or the storage device.
This way, they improve the WRENCH performance model for reading data.
Though, they do not present any results on the write performance, nor do
they cover the OS throttling behavior and the relevant
Linux parameters
to regulate the dirty pages in the page cache, such as
$\bgrate$, $\dirtyratio$, $\posratio$, etc.

The Linux kernel documentation~\cite{linuxdoc} contains
more details on the discussed parameters and also some hardware-specific
parameters of storage devices (queue length, buffer size, physical block size, etc.)
that define the raw performance behavior of a particular storage device (though
discussing the hardware-specific parameters is beyond the scope of our paper).

In the domain of storage device performance modeling, Huang~\cite{ssdmodel} models the performance of
SSD devices based on their hardware architecture and machine learning techniques.
As opposed to our model, Huang et al.~\cite{ssdmodel} do not consider the OS and library
caching mechanisms. Instead, the model describes specifically SSDs
at a lower level. Yagdar et al.~\cite{ssdcat} analyzed the IO workloads of SSDs and
the amplifying page size effects on the read and write operations. Though similar
to Huang's approach,
Yagdar's study is very specific to SSDs and dives into the
SSD hardware architecture.
Desnoyers~\cite{Desnoyers14} explores the write amplification of SSDs under random IO
considering their internal page management and models the performance of SSDs at a
lower level (SSD-internal) for hot and cold data.
Another effort to model SSD performance is done by Kim~\cite{KimCLK21} which similar to
our methodology, predicts the latency of the next access to the SSD considering key
performance-critical features.
IOSIG~\cite{iosig} is a GCC plugin that enables
the C/C++ software developers to pack their knowledge on the application’s IO
into the source code using C/C++ pragma notions
and in the form of a set of parameters. Then, IOSIG uses a cost model
to predict the IO throughput of the application considering the defined
parameters, and chooses the best-fitting storage device for the
application. This is analog to our work in the sense of
describing the IO cost using a model.
Though, also IOSIG does neither take the cache management of
the OS nor the C standard library buffer into account.
Another approach to characterize workloads of different patterns
if done in \cite{GuzLSB18}, where the block-level IO traces of billions of IO requests
from two large Cloud systems are analyzed for load balancing, cache-efficiency, and
storage cluster management. In this regard, our findings and measurements could be used
to analyze Cloud workloads to increase cache-efficiency and improve load balancing.

Lebre et al.~\cite{lebre2015adding} provide an extension for SimGrid to simulate storage
capabilities on distributed computing systems. They take the network
infrastructure of the storage system, device characteristics,
interconnections between processes, resource sharing, etc. into account
and model the storage performance of the distributed application as a whole.
CART models~\cite{cart} provide a black-box machine learning algorithm that
are trained with the gained performances of a storage device for different
workloads and learn the device's behavior without having any knowledge about the
device.

\section{Evaluation} \label{sec:eval}

We evaluate our model CAWL with regard to the actual IO costs in different experiments and compare the
accuracy of it with the
IO model of the SimGrid~\cite{simgrid} simulation framework. SimGrid estimates the IO costs
naively by dividing the data size to the bandwidth of the storage device. Hence, it disregards
the involved caches of the OS and the C standard library.
We perform experiments in different scenarios, on different hosts and storage devices.
Although, due to space limitations, we have to confine to a
subset of representative experiments on two hosts represented in \cref{tab:hosts}.
For each plot we present the average relative errors of the models ($\delta_{CAWL}$ and $\delta_{simgrid}$)
compared to the actual IO costs.
Moreover, \emph{Improvement} is defined for
each plot to demonstrate the estimation improvement achieved by CAWL compared to SimGrid. The durations of performing IO simulations (without the actual experiments) of CAWL and SimGrid on host 1 were
$\approx0.0005\,s$ for both of them in most of the cases.

\begin{table}
\caption{Hosts' Properties}
\label{tab:hosts}
\centering
\footnotesize
\begin{tabular}{r|r|r}
  & \textbf{Host 1: RAID 5 Host}\hfil\mbox{}
  & \textbf{Host 2: SSD Host}\hfil\mbox{} \\
  \hline    \hline
    CPU               & 4$\times$Intel Xeon Platinum 8353H           &2$\times$ Intel Xeon E5-2630v3\\
                      & CooperLake @ 2.50\,GHz                        &\\
    \hline
    RAM& 384\,GB (DDR4-3200 RDIMM)                     & 64\,GB\\
    \hline
    Storage         & 4$\times$ 5.8\,TB NVMe (Raid5)                &500 GB V-NAND SSD\\
    Device          & \emph{Model:}\hfill Intel SS PHLN12320               &\emph{Model:}\hfill Samsung 850 EVO\\

                    & \emph{Filesystem:}\hfill ext4                        &\emph{Filesystem:}\hfill ext4\\
                    & \emph{Logical sector size:} \hfill 512\,B  & \emph{Logical sector size:} \hfill 512\,B\\
    \hline
    OS              & CentOS Stream release 8\hfill\text{  }    & Debian 5.10.120-1\hfill\text{  }\\
                    & \emph{Kernel:}\hfill\text{  }          & \emph{Kernel:}\hfill\text{  }\\
                    & 4.18.0-394.el8.x86\_64 &\hfill 5.10.0-15-amd64\\
    \hline
    sysctl          & \emph{dirty\_background\_ratio:}\hfill 10\%             & \emph{dirty\_background\_ratio:}\hfill 10\%\\
                    & \emph{dirty\_ratio:}\hfill 40\,\%             & \emph{dirty\_ratio:}\hfill 20\,\%\\
                    & \emph{dirty\_expire\_centisecs:}\hfill 3000             & \emph{dirty\_expire\_centisecs:}\hfill 3000\\
\end{tabular}

\end{table}

\subsection{Benchmarking Tool}

To estimate the IO cost, our model requires different system parameters, as discussed in \cref{sec:model}.
Hence, we
developed a benchmarking tool in \Cpp that detects, fetches, or measures all the required
parameters in a system and stores them in a config file~\cite{cawlgit}.

Memory bandwidth ($\mbw$) is defined by measuring the costs of
copying large data chunks in the memory.
To measure $\rambw$, the benchmarking tool periodically issues
large IO requests---one after each other---until the rate of the dirty pages
reaches the $\bgrate$.
At this point the OS starts to sync in the background.
Hence, the $\redbw$ is defined by continuing to issue IO requests and measuring the increased IO costs.
To measure the device bandwidth and system call
costs, IO requests of two different size ranges are issued to the device
(using the \ODIRECT flag).
IO chunks with a range of small sizes (e.g., a few kilobytes) are performed repeatedly (to get a
reasonable average cost for each size), and the resulting costs
are analyzed using linear regression to fetch the system call cost
($\just{cost} = \just{size}/\just{sbw} + \sysswcost$). Theoretically, we could also get the device
bandwidth by the linear regression analysis. Though, to get a more consistent
average device bandwidth ($\just{sbw}$),
we perform more IO requests with a range of large sizes and
again analyze the IO costs using linear regression to get the device bandwidth.

The logical block size is fetched from
\emph{/sys/block/$<$sdX$>$/queue/logical\_block\_size}, and the dirty page
expire time from \emph{/proc/sys/vm/dirty\_expire\_centisecs}.
The $\dirtyratio$ and $\bgrate$ are read from \emph{/proc/vmstat}
(under the names \emph{nr\_dirty\_threshold} and \emph{nr\_dirty\_background\_threshold}). The buffer size of the C library is determined using
the builtin function \emph{\_\_fbufsize}~\cite{fbufsize}. After running
our benchmarking tool~\cite{cawlgit} on the two hosts, all required parameters are
determined automatically and stored to a config file.
Our model (also implemented in \Cpp~\cite{cawlgit}) uses the config file
to estimate the IO cost in different scenarios.

\subsection{Evaluating Synchronized and Direct IO with System Calls}

\begin{figure}
	\begin{subfigure}{0.49\linewidth}
    \centering%
\includegraphics[width=1\linewidth]{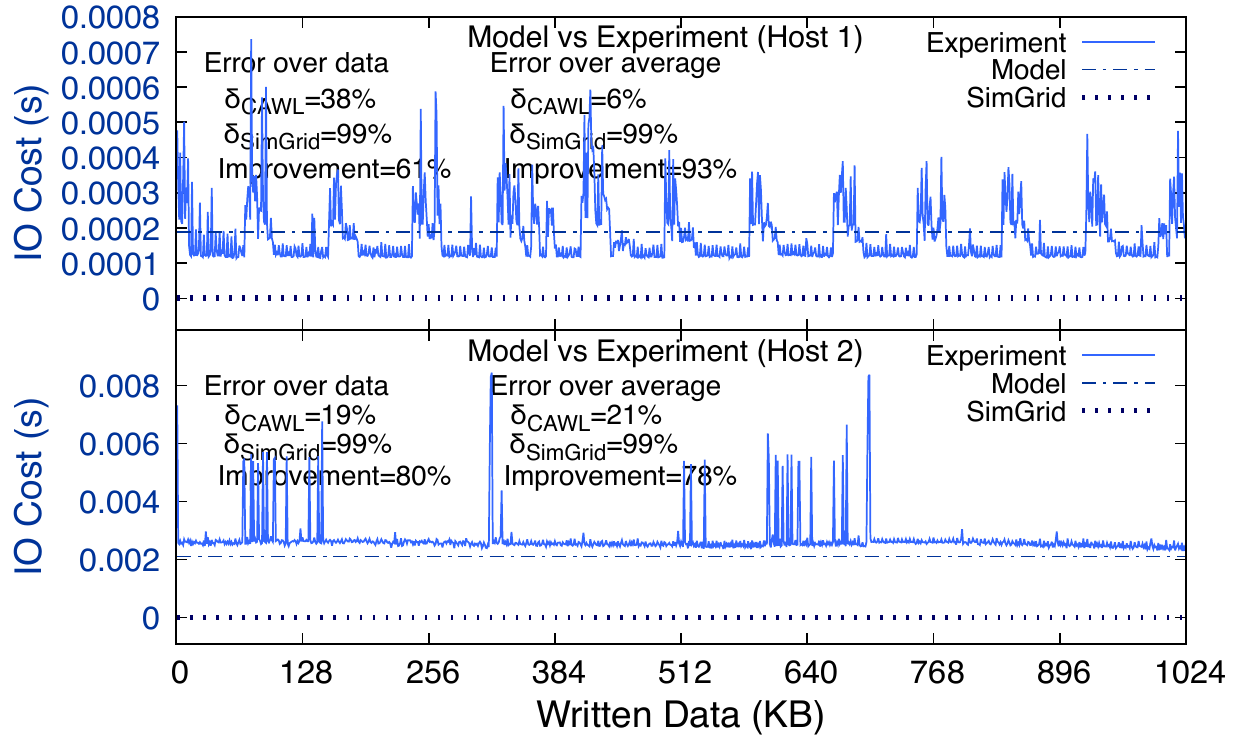}
\caption{Direct IO of $1\,\MB$ data in $1\,\KB$ chunks.
	\label{fig:gplot_direct_cost}}
	\end{subfigure}
	\begin{subfigure}{0.49\linewidth}
    \centering
    \includegraphics[width=1\linewidth]{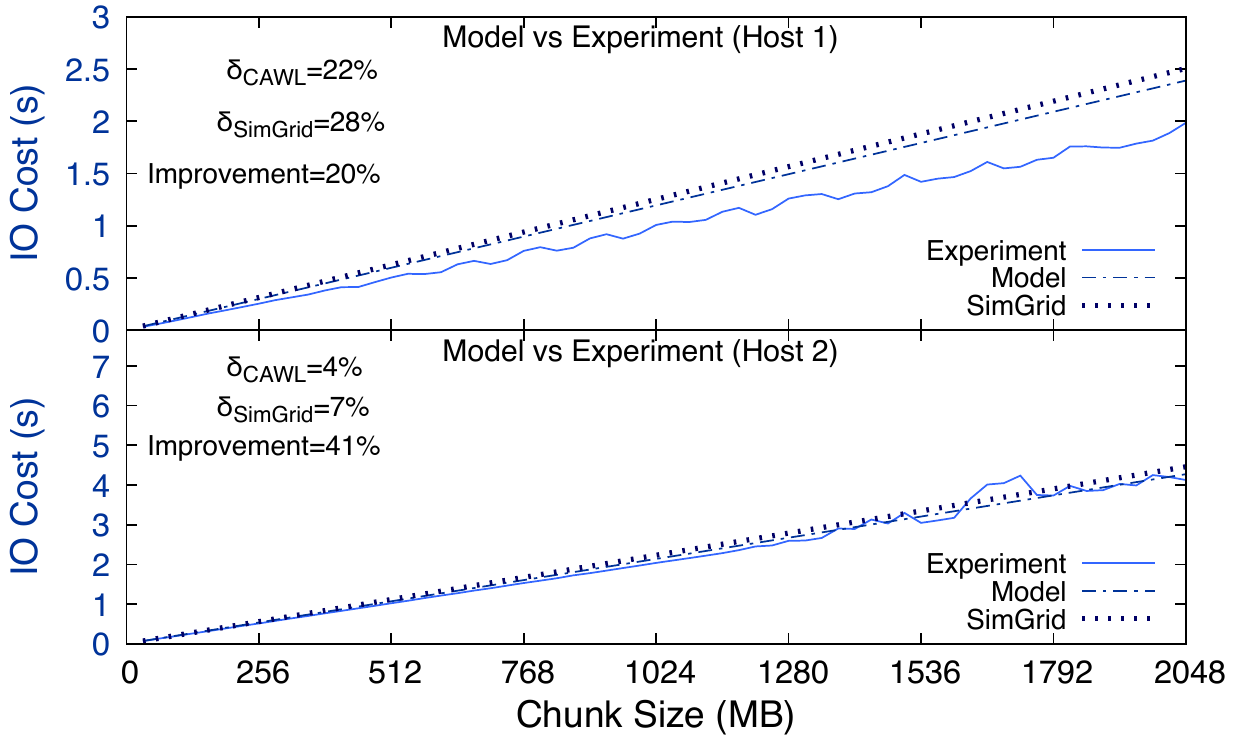}
    \caption{Direct IO for chunk sizes in the range $32\,\MB$--$2\,\GB$.\label{fig:direct_cost_range}}
  \end{subfigure}
  \caption{Model versus experiment: \emph{direct} IO for different amounts of
  written data and different chunk sizes.}
\end{figure}

In the case of direct IO, the OS does not come in between (see the
case 1 of \cref{fig:iopath}),
and the data goes directly to the storage device.
\Cref{fig:gplot_direct_cost}
compares the model predictions (CAWL and SimGrid) with the IO costs of writing $1\,024$ data chunks, each of size $1\,\KB$
(i.e., $1\,\MB$ data in total), on the two hosts with different
storage devices.
The IO costs on host 1 follow a repeating pattern of peaks and troughs.
The shape of this pattern is determined by
the specifications of the storage device's queue, such as
the maximum number of requests~\cite[nr\_requests]{devqueue}, maximum
segments and segment size~\cite[max\_segments, max\_segment\_size]{devqueue},
and other hardware/middleware specifications. Since discussing the
storage-device-specific parameters are beyond the scope of our work we make generic assumptions
in our model.
Because of this pattern, we also added the relative error (i.e., the absolute error divided by the actual value) of the estimations
compared to
the \emph{average} IO costs on the plot.
As you see, CAWL provides reasonably well generic
approximations of the overall performance of the storage devices with an average
relative errors of $\delta_{CAWL}=6\%\, \&\, 19\%$ on hosts 1 \& 2, while SimGrid
underestimates the IO costs with relative error
$\delta_{simgrid}=99\%$ on both hosts because it disregards the system call costs, which
are non-negligible in this case.
Additionally, \cref{fig:direct_cost_range} compares the measured
direct IO costs with the model for a range of IO chunks between
$32\,\MB$--$2\,\GB$ on the two hosts.

\begin{figure}
	\begin{subfigure}{0.49\linewidth}
    \includegraphics[width=1\linewidth]{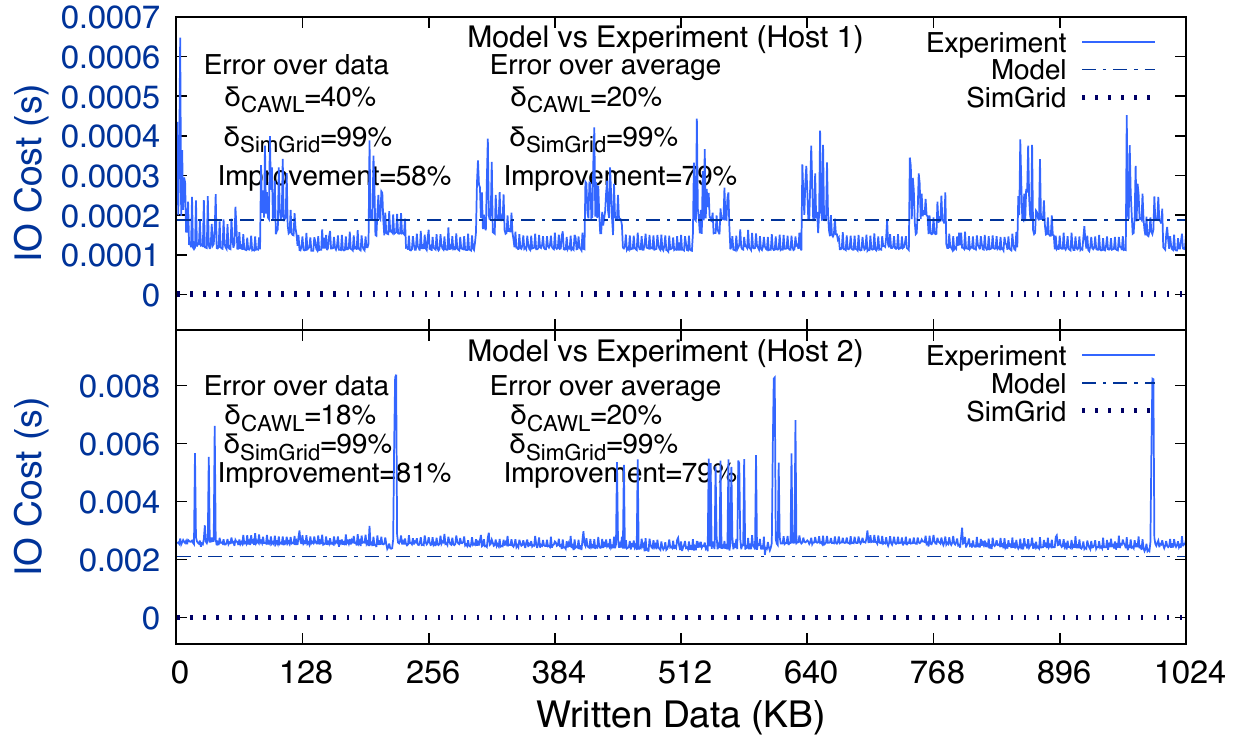}
\caption{Synchronized IO of $1\,\MB$ data in $1\,\KB$ chunks.
	\label{fig:gplot_sync_cost}}
	\end{subfigure}
	\begin{subfigure}{0.49\linewidth}
    \centering
\includegraphics[width=1\linewidth]{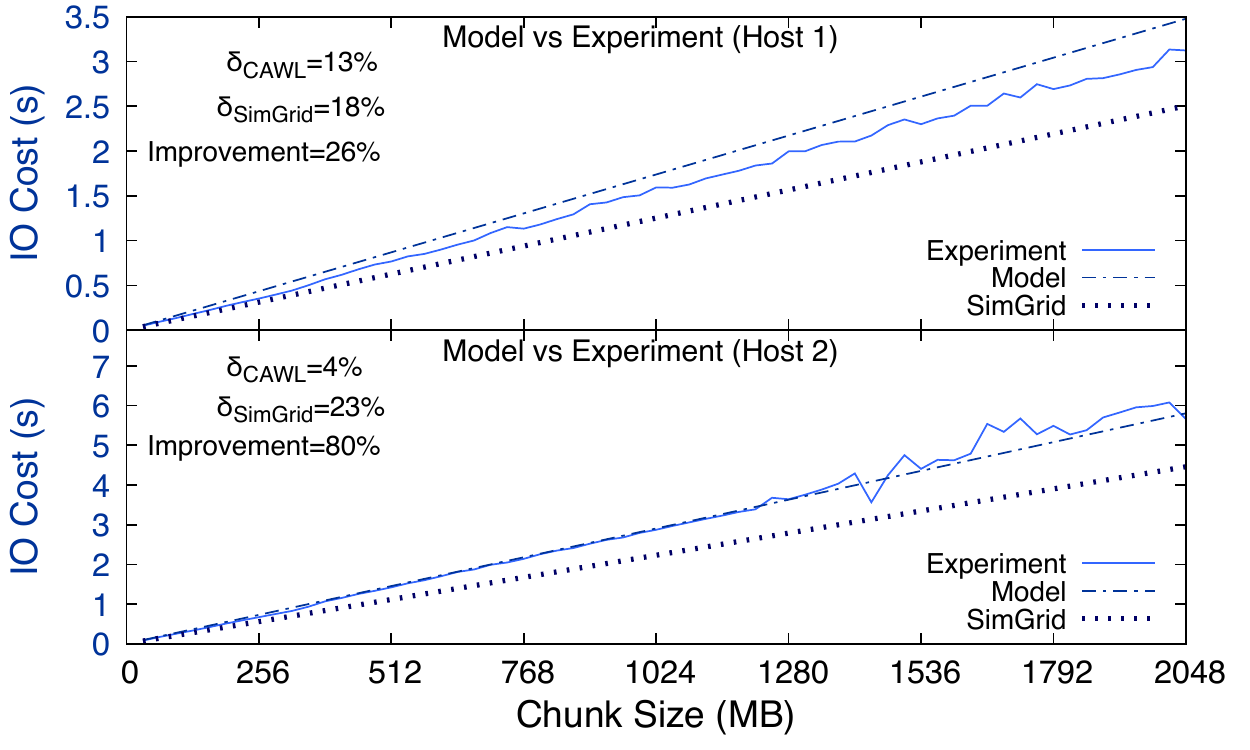}
\caption{Synchronized IO for chunk sizes of $32\,\MB$--$2\,\GB$.
	\label{fig:sync_cost_range}}
	\end{subfigure}
  \caption{Model versus experiment: \emph{synchronized} IO for different amounts of
  written data and chunk sizes.}
\end{figure}

\subsection{Evaluating Synchronized, Non-direct IO with System Calls}

\Cref{fig:gplot_sync_cost} compares the predictions of the models with the actual
IO costs of writing
$1\,024$ data chunks of size $1\,\KB$, with synchronized non-direct IO on the two
hosts.
Here, we have
the fixed extra cost of copying into the OS cache.
Hence,
the pattern of IO costs is similar to the direct IO
(\cref{fig:gplot_direct_cost}) with a minor extra overhead.
The copying overhead is negligible compared to the system call cost and
hence is hardly visible on \cref{fig:gplot_sync_cost}. But
it is observable in \cref{fig:sync_cost_range}, which compares the models
with the costs of writing data chunks of different sizes
between $32\,\MB$--$2\,\GB$ on
the two hosts. As you see, the SimGrid model fails to follow along the actual
IO costs (relative error $\delta_{simgrid}=18\%\,\&\,23\%$) since it does not account for the copying overhead.
On the other hand, CAWL takes this effect into account and predicts the IO costs
with a relative error $\delta_{CAWL}=13\%\,\&\,4\%$ (an overall
estimation improvements of $26\%\,\&\,80\%$ respectively).

\subsection{Evaluating Non-Synchronized IO with System Calls}

\begin{figure}
    \centering
        \begin{subfigure}{0.49\linewidth}
            \includegraphics[width=1\linewidth]{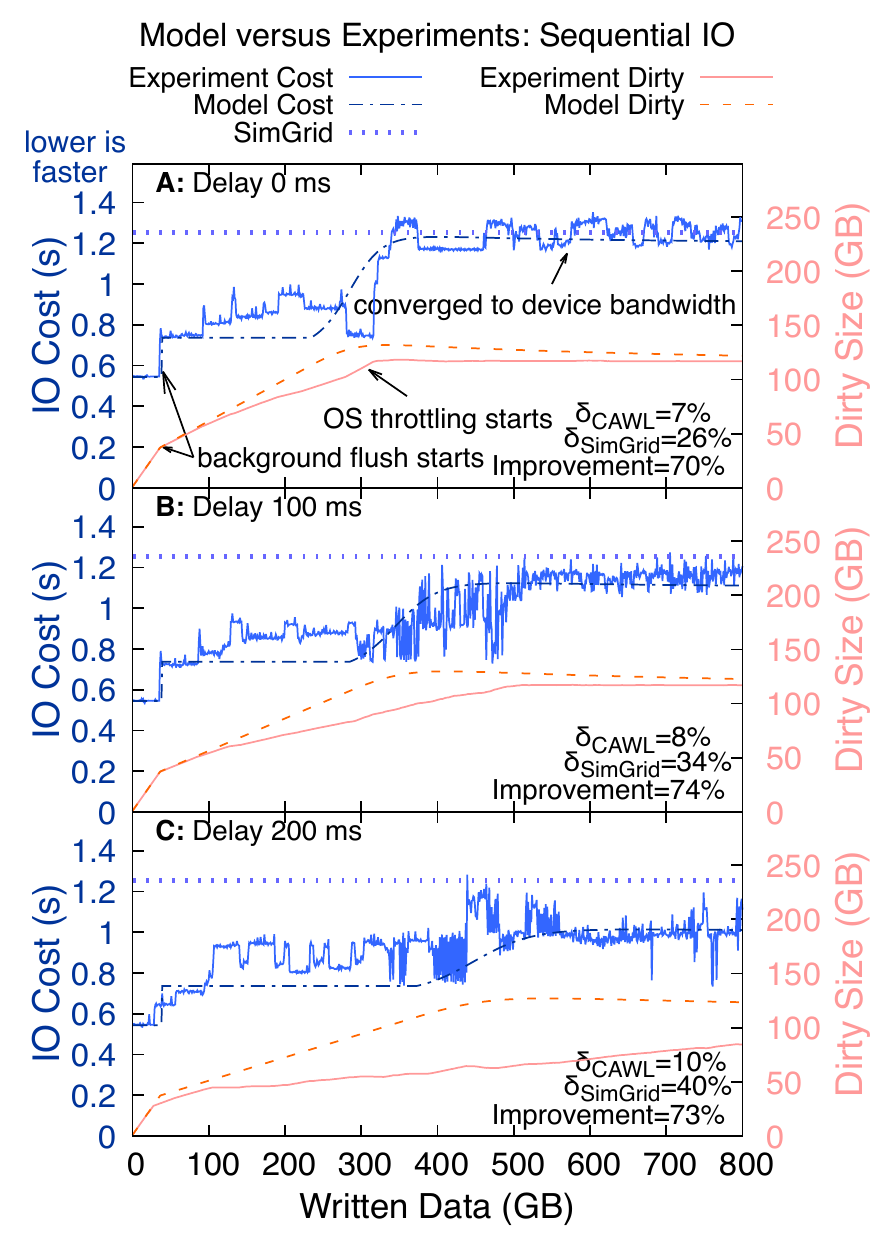}
            \caption{Host 1}
        \end{subfigure}
        \begin{subfigure}{0.49\linewidth}
            \includegraphics[width=1\linewidth]{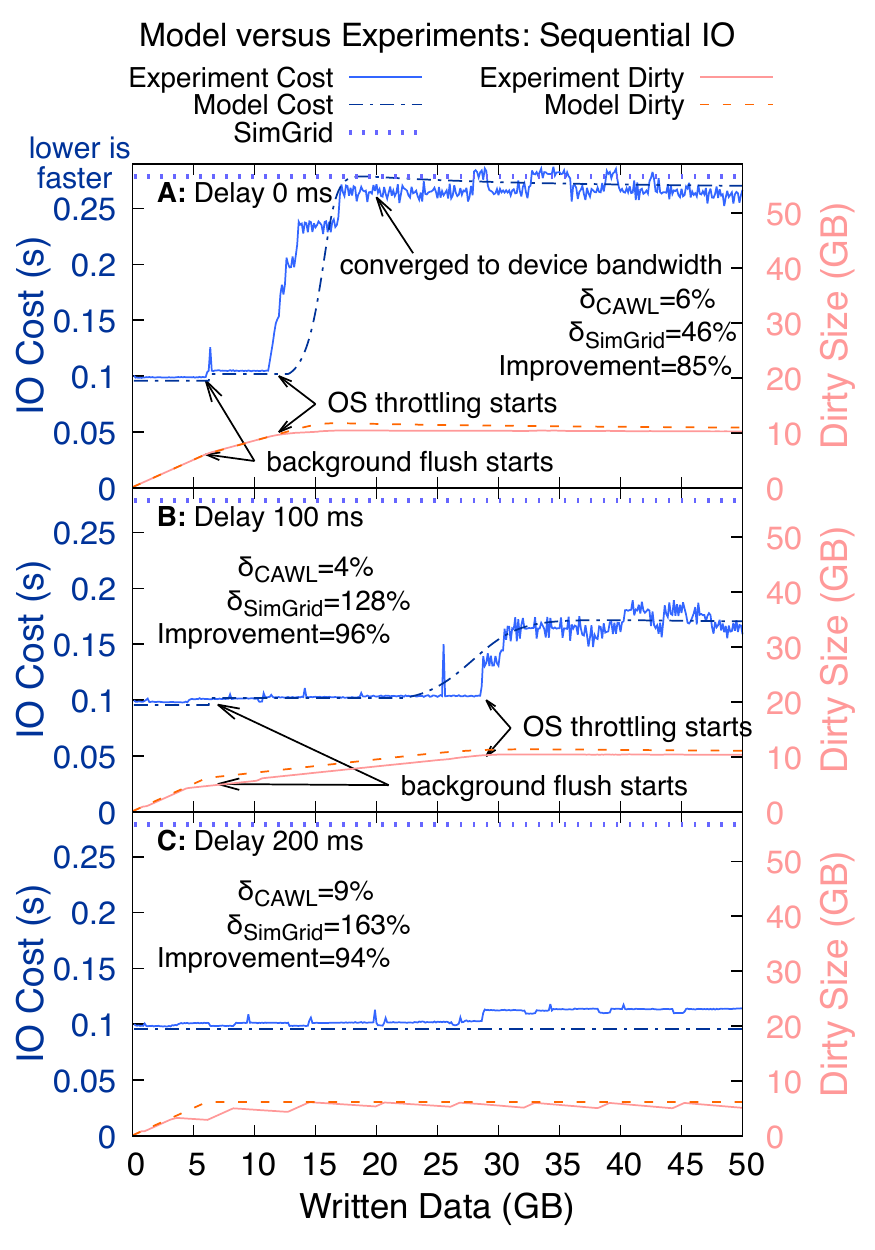}
            \caption{Host 2}
        \end{subfigure}
    \caption{Non-synchronized IO, chunk size $1\,\GB$.\label{fig:async_cost}}
\end{figure}

\begin{figure}
    \centering
        \begin{subfigure}{0.49\linewidth}
            \includegraphics[width=1\linewidth]{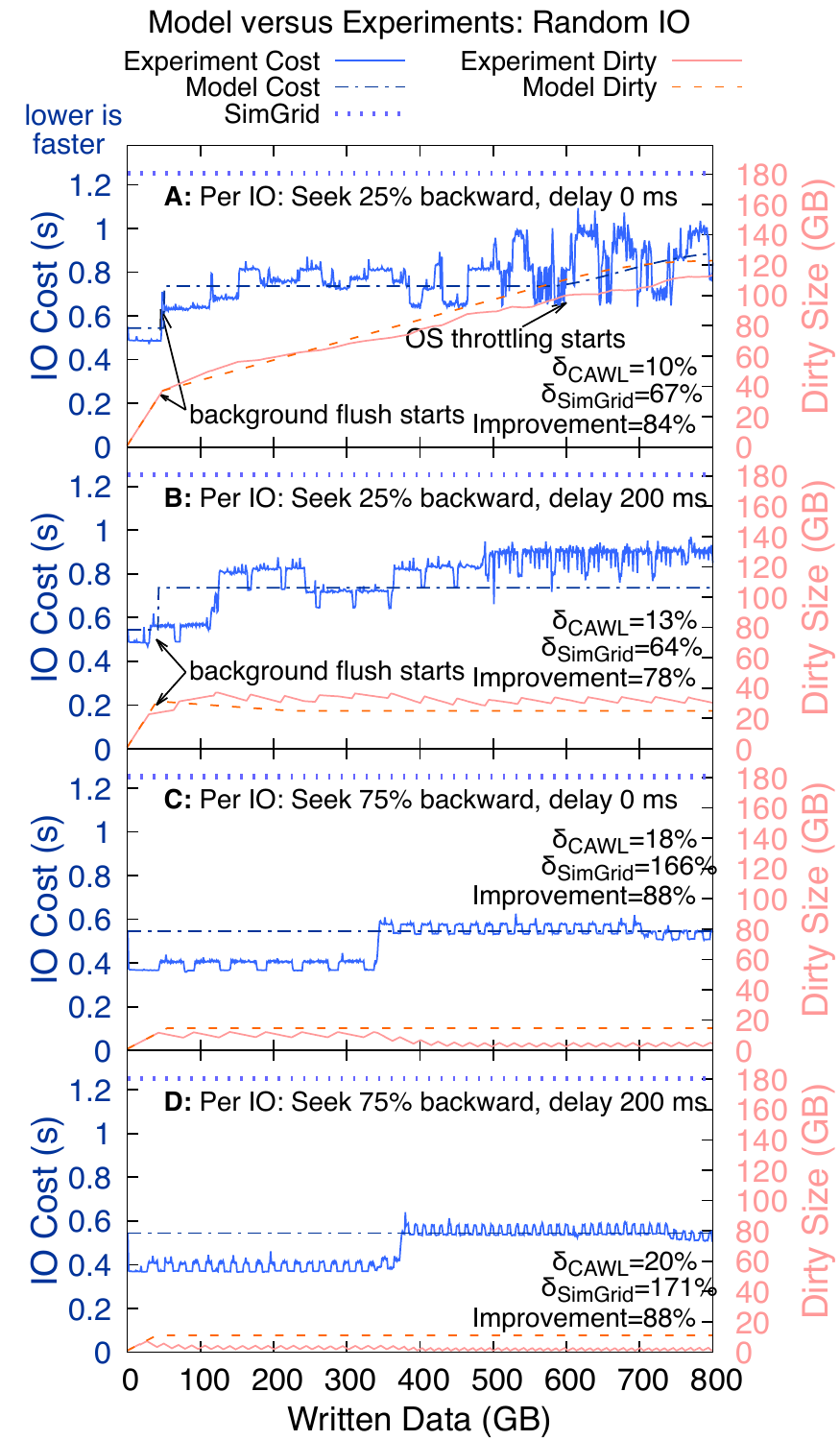}
            \caption{Random IO on host 1\label{fig:async_random_host1}.}
        \end{subfigure}
        \begin{subfigure}{0.49\linewidth}
            \includegraphics[width=1\linewidth]{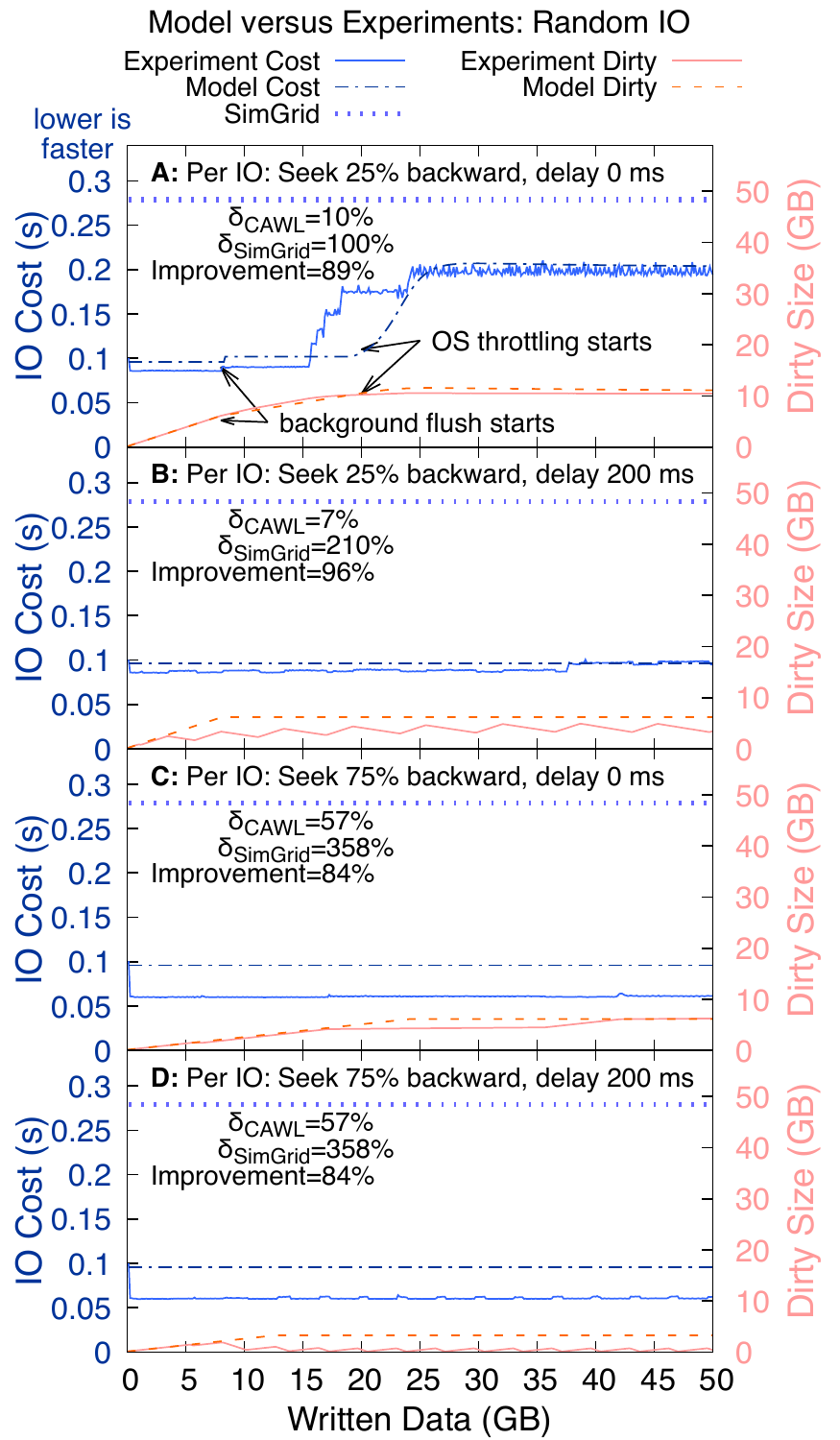}
            \caption{Random IO on host 2\label{fig:async_random_host2}.}
        \end{subfigure}
    \caption{Random IO with non-synchronized system calls (chunk size $1\,\GB$).\label{fig:async_random}}
\end{figure}

For non-synchronized IO with system calls, the OS caches the data before writing it to the storage
device.
\Cref{sec:mod-nonsync} discussed the effects of $\bgrate$ and $\dirtyratio$.
Here, we evaluate the model in two cases of $\dirtyratio=40\,\%$
(host 1) and $\dirtyratio = 20\,\%$ (host 2).
\Cref{fig:async_cost}-A compares the measured
non-sync IO costs with the model values of CAWL and SimGrid for
writing $800\,\GB$ and $50\,\GB$ data in $1\,\GB$ and $128\,\MB$
chunks sequentially on hosts 1 (left) and 2 (right), respectively.
As expected, SimGrid cannot appropriately predict the IO cost with the OS cache
involved, constantly estimating the performance of the storage device.
In this case, SimGrid overestimates the IO costs, as opposed to the previous cases
where it underestimated the IO costs.
CAWL on the other hand, reasonably models the IO costs and smoothly follows
the actual costs. Moreover, CAWL correctly predicts
the points where the background flushing and throttling of the OS starts
(annotated on the plots).
CAWL also considers the computations between IO operations.
The right-hand y-axis of plots represents the actual amount of dirty pages in the system, as well as the approximation of the dirty amount by CAWL.
\Cref{fig:async_cost}-B and \cref{fig:async_cost}-C
show the comparison between the experiments and the model values
for writing $800\,\GB$ and $50\,\GB$ data in $1\,\GB$ and $128\,\MB$
chunks sequentially (on hosts 1 (left) and 2 (right), respectively) and computing for
$100\,ms$ and $200\,ms$ (hashing random data) before writing each chunk.
In this case, CAWL takes the ongoing background flush during the computation
into account and still estimates the IO costs reasonably well.

CAWL does not only take the computations before IO into account, but also considers the randomness of IO.
\Cref{fig:async_random} compares the values of CAWL, SimGrid, and the experimental
results in the case of random IO when having the following
situations before writing each subsequent data chunk:
\begin{enumerate}[label=(\Alph*)]
    \item no computation (delay = $0\,ms$), seek backward to rewrite the last
        $256\,\MB$ of the written data (rewrite $25\,\%$ of chunk size). Hence, $75\,\%$
        of each data chunk is new data.
    \item compute for $200\,ms$, seek backward to rewrite the last
        $256\,\MB$ of the written data.
    \item no computation, seek backward to rewrite the last
        $768\,\MB$ of the written data (rewrite $75\%$ of chunk size).
    \item compute for $200\,ms$, seek backward to rewrite the last
        $768\,\MB$ of the written data.
\end{enumerate}

Considering the relative errors of both models compared to the actual IO costs,
overall, the results
suggest that CAWL usually estimates the IO cost with a relative error
less than $10-20\,\%$, which is $70\,\%$ to $90\,\%$ more accurate
than SimGrid.

\subsection{Modeling Library IO}

\begin{figure}
	\centering
		\begin{subfigure}{0.49\linewidth}
			\includegraphics[width=1\linewidth]{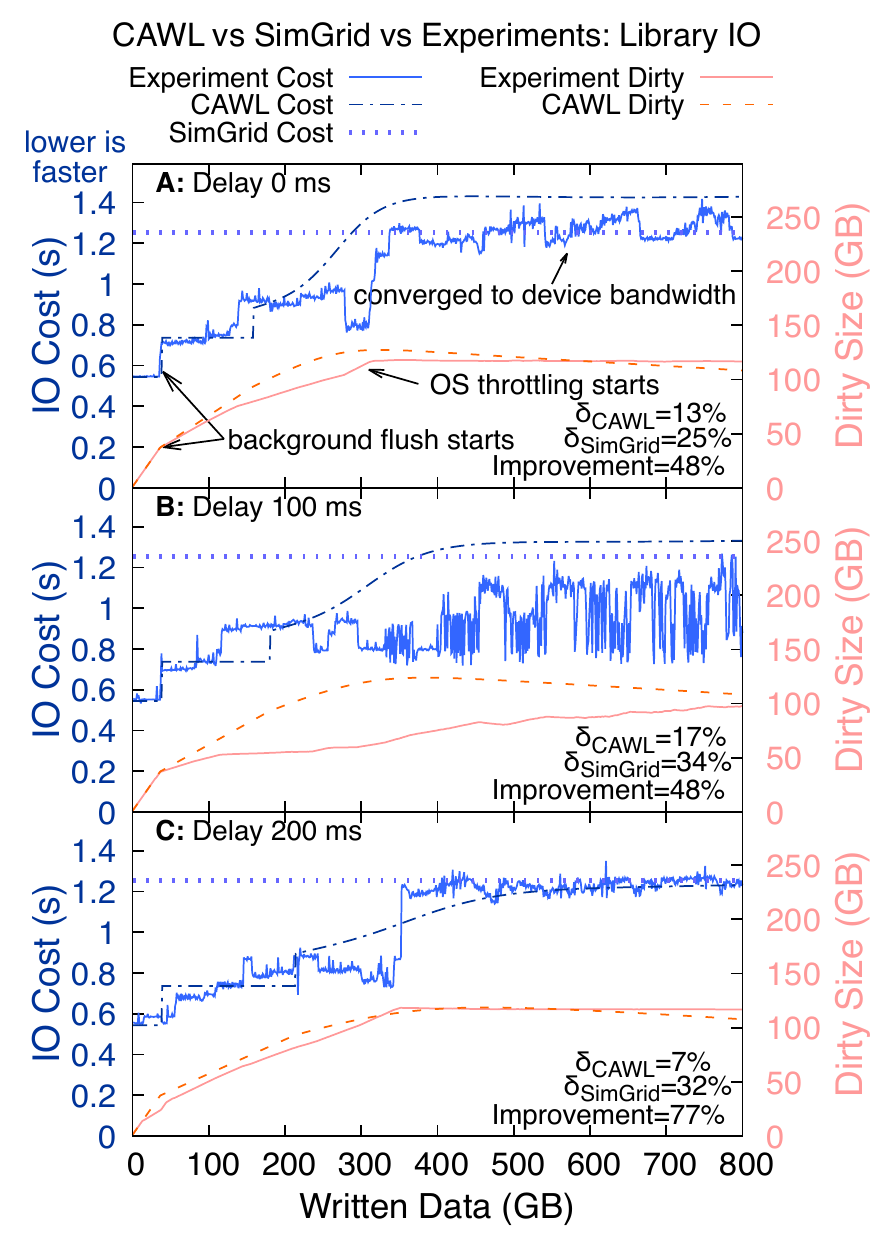}
			\caption{Host 1}
		\end{subfigure}
		\begin{subfigure}{0.49\linewidth}
			\includegraphics[width=1\linewidth]{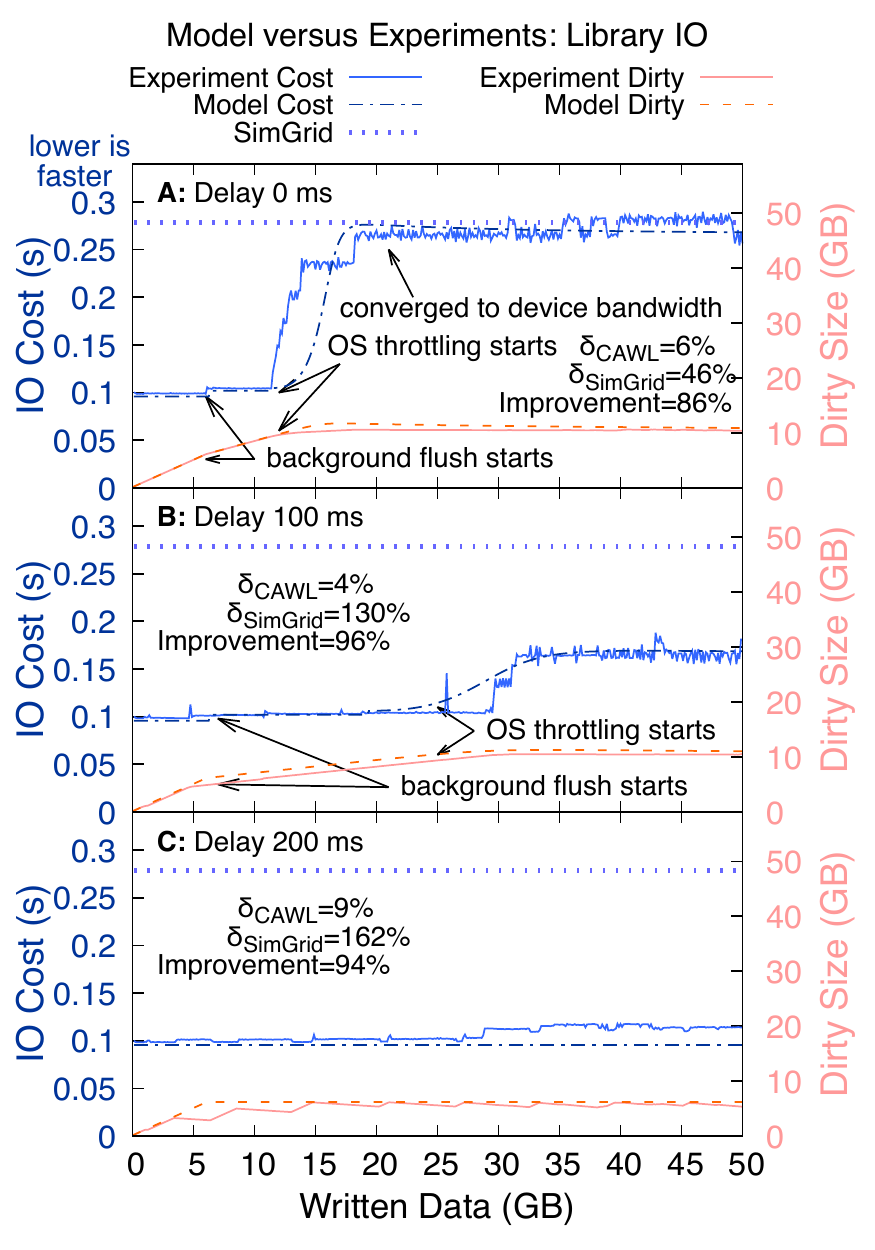}
			\caption{Host 2}
		\end{subfigure}
	\caption{C standard library IO, chunk size $1\,\GB$.\label{fig:lib_cost}}
\end{figure}

\begin{figure}
	\includegraphics[width=0.5\linewidth]{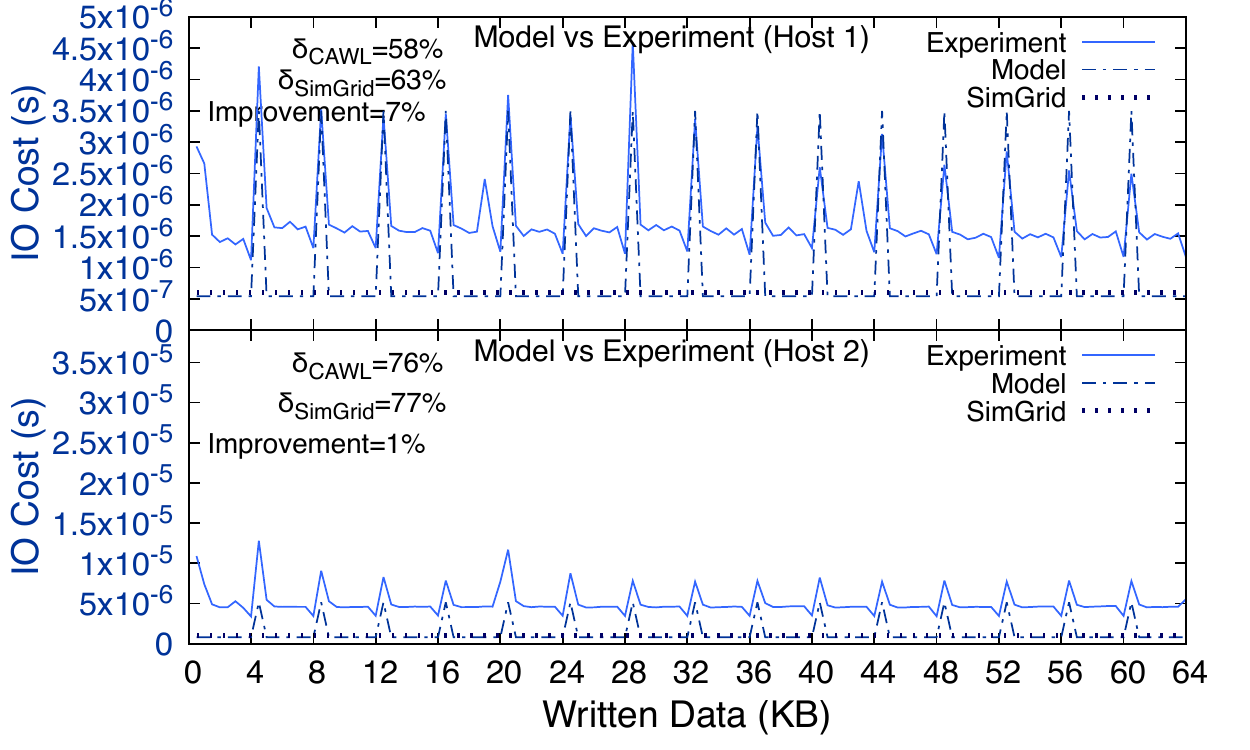}
	\caption{C standard library IO: effects of the buffer size.
		\label{fig:lib_buffer}}
\end{figure}

Lastly, \cref{fig:lib_cost}
compares the measured IO costs with the estimations of CAWL and SimGrid when
using C standard library IO to write $800\,\GB$ and $50\,\GB$
(on the hosts 1 (left) and 2 (right), respectively) in $1\,\GB$ and $128\,\MB$ chunks and
delays of $0\,ms$ (no delays), $100\,ms$, and $200\,ms$.
In addition, we discussed the buffer of the C standard library and considered
its effect in CAWL. \Cref{fig:lib_buffer} illustrates this effect in the
case of writing $64\,\KB$ of data in $512\,B$ chunks.

\section{Conclusion} \label{sec:conc}

We presented CAWL, a model designed to assess the IO costs by taking into account the Linux kernel page cache and the C standard library IO buffer. Our model effectively captures the IO costs of various types of operations, including sequential and random IO, with and without computational tasks interspersed, across multiple classes of IO operations. By comparing our model with the SimGrid IO model, we demonstrated that the latter can suffer from significant underestimation or overestimation of IO costs, particularly in IO-intensive applications due to its simple IO model that does not consider caches and buffers. As a plug-in, CAWL could complement SimGrid's capability in this area.

Furthermore, our model achieved a high level of accuracy, accurately predicting IO costs at around 80--90\,\%, while considering different scenarios and the presence of caches. These findings, along with our measurements of Linux page cache management, have practical implications in both Cloud environments, where they can enhance cache efficiency and load balancing, and in high-performance computing (HPC) environments, where they can be utilized to simulate the lifespan of workflows.

The model parameters of CAWL and their influence could also provide valuable insights for hardware design. By considering various scenarios and caches, CAWL's accurate estimations of IO costs can guide cache design, optimize storage systems, and facilitate system-level performance analysis. Hardware designers could leverage CAWL's model parameters to make informed decisions with a better understanding of practical IO costs, leading to efficient resource allocation, reduced latency, and enhanced performance of their storage architectures.

You can find the implementation of our model, as well as our benchmarking tool that detects different system parameters required by the model on Github~\cite{cawlgit}.

\section*{Acknowledgment}
This work received funding from the German Ministry for Education and
Research as BIFOLD--Berlin Institute for the Foundations of Learning
and Data (ref. 01IS18025C) and the German Research Foundation (DFG)
as part of CRC 1404 FONDA. We thank ZIB's core facility and NHR@ZIB
units for providing compute resources.

\bibliographystyle{IEEEtran}

\bibliography{main}

\end{document}